\documentclass[aps,prb,reprint,twocolumn,nofootinbib,floatfix,superscriptaddress]{revtex4-2}

\usepackage{amsmath,amssymb,mathtools}
\usepackage{mathrsfs}
\usepackage{bm}
\usepackage{microtype}
\usepackage[colorlinks,linkcolor=blue,citecolor=blue,urlcolor=blue]{hyperref}

\newcommand{\Tr}{\operatorname{Tr}}

\begin{document}

\title{Dissipative many-body theory of electron-phonon-photon systems}

\author{Zhenlin Zhang}
\affiliation{Hefei National Research Center for Physical Sciences at the Microscale, University of Science and Technology of China, Hefei, Anhui 230026, China}

\author{Enrico Perfetto}
\affiliation{Dipartimento di Fisica, Universit\`a di Roma Tor Vergata, Via della Ricerca Scientifica 1, 00133 Rome, Italy}
\affiliation{INFN, Sezione di Roma Tor Vergata, Via della Ricerca Scientifica 1, 00133 Rome, Italy}

\author{Gianluca Stefanucci}
\email{gianluca.stefanucci@roma2.infn.it}
\affiliation{Dipartimento di Fisica, Universit\`a di Roma Tor Vergata, Via della Ricerca Scientifica 1, 00133 Rome, Italy}
\affiliation{INFN, Sezione di Roma Tor Vergata, Via della Ricerca Scientifica 1, 00133 Rome, Italy}

\date{\today}

\begin{abstract}
Optical cavities offer opportunities to control quantum materials, but
radiative leakage and vibrational relaxation call for microscopic theories
that treat interactions and dissipation on equal footing. We develop a
Keldysh-Lindblad nonequilibrium Green's function framework for interacting
electrons coupled to quantized photon and phonon fields. The formulation
accommodates boson-number-nonconserving Hamiltonians and dissipative channels
that combine electronic transitions with bosonic processes. We derive the
mixed electron-boson diagrammatic expansion and the corresponding
Kadanoff-Baym equations, retaining environmental noise, initial correlations,
and interaction-induced memory. The equal-time equations govern the electronic
density matrix and the bosonic covariance. As an illustration, we construct
jump operators that preserve a semiconductor with filled valence bands and
an empty photon field. Within a mean-field treatment, this state remains
stationary although its two-time propagators exhibit dissipative broadening.
The framework opens new avenues for studying correlated dynamics in
lossy cavities and vibrational environments.
\end{abstract}

\maketitle

\section{Introduction}
\label{sec:introduction}

Embedding quantum materials in optical cavities offers new ways to control
collective excitations and electronic properties through the quantized
electromagnetic field~\cite{GarciaVidal2021VacuumFields,
Schlawin2022CavityQuantumMaterials,Basov2016VdWPolaritons}.
Atomically thin semiconductors are especially promising because their
strong excitonic resonances and valley-selective optical transitions
allow light and matter to hybridize efficiently~\cite{Wang2018TMDExcitons,
MakShan2016TMDPhotonics,MuellerMalic2018TMDExciton}.
Strong coupling in semiconductor monolayers and van der Waals
heterostructures has been demonstrated experimentally, and cavity
control of excitons and vibrational degrees of freedom has been
explored theoretically~\cite{Liu2015StrongCoupling2D,
Dufferwiel2015VdWCavity,Flatten2016WS2Polaritons,
Latini2019CavityExcitons,Sentef2018CavityEph,Latini2021Phonoritons}.
In practice, however, electromagnetic confinement is accompanied by
radiative leakage and absorption, while material excitations undergo
scattering and decoherence~\cite{Selig2016TMDLinewidth,
Christiansen2017PhononSidebands,ScheelBuhmann2008MacroscopicQED,
Lalanne2018,Franke2019QNM}.
Understanding cavity-modified materials therefore requires a theory
that describes their microscopic many-body structure together with
dissipation.

A large part of cavity quantum electrodynamics relies on effective
models inherited from quantum optics, in which matter is represented
by a few electronic levels or collective excitations coupled to a
small number of photon modes~\cite{Schlawin2022CavityQuantumMaterials}.
Such models clarify the competition between coherent coupling and
loss, but their material parameters are generally prescribed rather
than derived, and systematic descriptions of electronic bands,
correlations, and electron-phonon coupling are difficult to obtain.
Quantum-electrodynamical density-functional theory (QEDFT) provides a
microscopic alternative by treating electrons and quantized photon
fields within an electronic-structure framework
~\cite{Tokatly2013TDQEDFT,Ruggenthaler2014QEDFT,
Flick2015QEDFT,Ruggenthaler2023AbInitioQED}.
Most commonly used cavity-QEDFT formulations employ discrete,
lossless photon modes. Extensions to continuous photonic spectra and
dissipative cavities have also been developed
~\cite{Kudlis2022DissipativeQEDFT,Flick2022DissipativeQEDFT},
but a general diagrammatic treatment of correlated electron-photon
and electron-phonon dynamics in an explicitly dissipative setting
remains desirable.

This need is particularly acute for lossy cavities, where dissipative processes affect not only the widths of spectral resonances but also the evolution of electronic and bosonic correlations. Introducing phenomenological linewidths into an effective Hamiltonian does not, by itself, account for quantum jumps and their effects on interacting particles. Furthermore, microscopic light-matter Hamiltonians, such as the Pauli-Fierz Hamiltonian \cite{Ruggenthaler2014,Ruggenthaler2023}, and electron-phonon Hamiltonians \cite{stefanucci2023and,Giustino2017} generally contain terms that do not conserve photon or phonon number. A consistent many-body description must accommodate these number-nonconserving bosonic couplings alongside the dissipative channels of an open system.

Nonequilibrium Green's function (NEGF) theory provides a natural
framework for formulating the many-body theory of such systems in and out of equilibrium. The
 Kadanoff-Baym equations (KBEs) describe correlated
dynamics on the Keldysh contour, including initial correlations and
memory effects~\cite{KadanoffBaym1962,Keldysh1965,
StefanucciVanLeeuwen2025}.
Its recent Keldysh-Lindblad extension incorporates Markovian
quantum jumps into second-quantized diagrammatic perturbation
theory~\cite{stefanucci2024kadanoff}.
This framework has enabled a first-principles non-Hermitian
Bethe-Salpeter equation for excitons coupled to dissipative
electromagnetic environments and systematic perturbative treatments
of dissipative quasiparticle dynamics
~\cite{zhang2026nhbse,Blommel2026DissipativeMBPT}.
Despite these advances, a corresponding unified diagrammatic
formulation for coupled matter and explicit bosonic fields, with
boson-number-nonconserving Hamiltonians and Lindblad operators that
mix matter and bosonic degrees of freedom, is still lacking.

In this work, we extend the Keldysh-Lindblad Green's function
formalism to interacting matter particles coupled to photon and
phonon fields, collectively referred to as \emph{phosons}.
The theory accommodates nonorthogonal single-particle bases and
general invertible transformations of doubled phoson fields.
We consider particle loss and gain, particle-bilinear and
phoson-linear jump operators, as well as shared jumps that generate
mixed dissipative interactions. We develop the corresponding
NEGF equations and diagrammatic
self-energies, and derive the KBEs for real-time, mixed, and Matsubara components.
Their equal-time limits yield coupled equations for the matter
density matrix and phoson covariance. Finally, we illustrate the formalism
with a semiconductor model whose filled valence bands and photon
vacuum remain stationary under the specified jump operators.
The resulting framework provides a starting point for systematic
many-body approximations to the dynamics and spectra of materials
in lossy photonic and vibrational environments.

The paper is organized as follows. In Sec.~\ref{The open electron-phonon-photon system}, we introduce the
interacting electron-phoson Hamiltonian and the Lindblad dynamics of the
open system. Section~\ref{Dissipative Hamiltonian on the Keldysh contour} recasts the dissipative evolution on the Keldysh
contour. In Sec.~\ref{Dissipative NEGF Formalism}, we establish the mixed Green's function formalism and
the corresponding diagrammatic expansion. Section~\ref{Kadanoff-Baym equations} presents the phoson KBEs in different Keldysh components. Section~\ref{Stationarity-preserving jump operators} illustrates the stationary
solutions of KBEs in a semiconductor model. Finally, Sec.~\ref{sec:conclusions} summarizes our
conclusions.

\section{The open electron-phoson system}
\label{The open electron-phonon-photon system}

\subsection{Closed system Hamiltonian in general basis}
Let $\hat\psi(\mathbf x)$, with $\mathbf x=(\mathbf r,\sigma)$,
annihilate an electron at position $\mathbf r$ with spin $\sigma$.
The field operators satisfy the canonical anticommutation relation
\begin{align}
    \{\hat\psi(\mathbf x),\hat\psi^\dagger(\mathbf x')\}
    &=\delta(\mathbf r-\mathbf r')\delta_{\sigma\sigma'}
    \equiv \delta(\mathbf x,\mathbf x').
\end{align}

We use the collective term \emph{phoson} to refer to photon- and
phonon-type bosons. Let $\hat a_\gamma$ and $\hat b_q$ denote
canonical photon and phonon annihilation
operators~\cite{Ruggenthaler2023AbInitioQED,Giustino2017ElectronPhonon},
satisfying
$[\hat a_\gamma,\hat a_{\gamma'}^\dagger]
=\delta_{\gamma\gamma'}$ and
$[\hat b_q,\hat b_{q'}^\dagger]=\delta_{qq'}$.

Let $N_a$ and $N_b$ denote the numbers of retained photon and
phonon modes, respectively, and introduce the column vectors
$\hat{\bm a}=(\hat a_1,\ldots,\hat a_{N_a})^{\mathsf T}$
and
$\hat{\bm b}=(\hat b_1,\ldots,\hat b_{N_b})^{\mathsf T}$.
The doubled phoson field is defined as
\begin{align}
    \hat\phi'_\nu
    =\begin{pmatrix}
        \hat{\bm a}\\
        (\hat{\bm a}^\dagger)^{\mathsf T}\\
        \hat{\bm b}\\
        (\hat{\bm b}^\dagger)^{\mathsf T}
    \end{pmatrix}_\nu.
    \label{eq:original-phoson-field-main}
\end{align}
Its commutation and Hermitian-conjugation relations read
$[\hat\phi'_\nu,\hat\phi'_{\nu'}] =\tilde\alpha_{\nu\nu'}$
and
$\hat\phi_\nu'^\dagger
=\sum_{\nu'}\tilde\kappa_{\nu\nu'}\hat\phi'_{\nu'}$,
where, in the ordering of
Eq.~\eqref{eq:original-phoson-field-main},
\begin{align}
    \tilde\alpha &=
    \begin{pmatrix}
        0&I&0&0\\
        -I&0&0&0\\
        0&0&0&I\\
        0&0&-I&0
    \end{pmatrix},
    &
    \tilde\kappa &=
    \begin{pmatrix}
        0&I&0&0\\
        I&0&0&0\\
        0&0&0&I\\
        0&0&I&0
    \end{pmatrix}.
    \label{eq:original-phoson-matrices-main}
\end{align}
The identity blocks in the photon and phonon sectors have
dimensions $N_a$ and $N_b$, respectively.
These matrices satisfy
$\tilde\alpha^{\mathsf T}=-\tilde\alpha$,
$\tilde\alpha^{-1}=-\tilde\alpha$, and
$\tilde\kappa^*\tilde\kappa=I$.

For the electronic degrees of freedom, we introduce a complete,
linearly independent set of basis functions
$\{U_n(\mathbf x)\}$, which need not be orthonormal.
The overlap matrix and dual basis functions are defined by
\begin{align}
    \mathcal M_{mn}
    &\equiv
    \int d\mathbf x\,U_m^*(\mathbf x)U_n(\mathbf x),
    \label{eq:electron-metric-main}\\
    \bar U_m(\mathbf x)
    &\equiv
    \sum_n(\mathcal M^{-1})_{mn}U_n^*(\mathbf x),
    \label{eq:electron-dual-main}
\end{align}
where
$\int d\mathbf x\equiv\sum_\sigma\int d\mathbf r$
and the overlap matrix $\mathcal M$ is assumed to be invertible.
The dual basis satisfies the biorthogonality relation
$\int d\mathbf x\,\bar U_m(\mathbf x)U_n(\mathbf x)
=\delta_{mn}$.

The corresponding electronic field operators are
\begin{align}
    \hat d_m
    &\equiv
    \int d\mathbf x\,\bar U_m(\mathbf x)\hat\psi(\mathbf x),
    \\
    \hat{\bar d}_m^\dagger
    &\equiv
    \int d\mathbf x\,\hat\psi^\dagger(\mathbf x)U_m(\mathbf x).
\end{align}
They satisfy
$\{\hat d_m,\hat{\bar d}_n^\dagger\}=\delta_{mn}$.
In a nonorthogonal basis,
$\hat{\bar d}_m^\dagger$ is generally not the
Hermitian conjugate of $\hat d_m$.
Instead, the two sets of operators are related by
\begin{align}
    \hat d_m^\dagger
    &=\sum_n\hat{\bar d}_n^\dagger
    (\mathcal M^{-1})_{nm},
    &
    \hat{\bar d}_m^\dagger
    &=\sum_n\hat d_n^\dagger\mathcal M_{nm}.
\end{align}

For the phoson degrees of freedom, we allow an arbitrary
invertible linear transformation of the doubled field,
\begin{align}
    \hat\phi_\mu
    =\sum_\nu S_{\mu\nu}\hat\phi'_\nu,
\end{align}
where $S$ is an invertible matrix.
The transformed field satisfies
\begin{align}
    [\hat\phi_\mu,\hat\phi_{\mu'}]
    &=\alpha_{\mu\mu'},
    &
    \hat\phi_\mu^\dagger
    &=\sum_{\mu'}\kappa_{\mu\mu'}\hat\phi_{\mu'},
\end{align}
with
\begin{align}
    \alpha
    &=S\tilde\alpha S^{\mathsf T},
    &
    \kappa
    &=S^*\tilde\kappa S^{-1}.
\end{align}
In particular,
$\alpha^{\mathsf T}=-\alpha$ and
$\kappa^*\kappa=I$.


In this general representation, the Hamiltonian governing the
coherent dynamics of the open electron-phoson system
takes the form
\begin{align}
    \hat H(t)
    &=\sum_{mn}h_{mn}(t)
    \hat{\bar d}_m^\dagger\hat d_n
    \nonumber\\
    &\quad+\frac{1}{2}\sum_{ijmn}v^{\rm phys}_{ijmn}(t)
    \hat{\bar d}_i^\dagger\hat{\bar d}_j^\dagger
    \hat d_m\hat d_n
    \nonumber\\
    &\quad+\frac{1}{2}\sum_{\mu_1\mu_2}
    Q_{\mu_1\mu_2}(t)
    \hat\phi_{\mu_1}\hat\phi_{\mu_2}
    \nonumber\\
    &\quad+\sum_\mu l_\mu(t)\hat\phi_\mu
    \nonumber\\
    &\quad+\sum_{\mu mn}g^{\rm phys}_{\mu,mn}(t)
    \hat\phi_\mu\hat{\bar d}_m^\dagger\hat d_n.
    \label{eq:model-H-main}
\end{align}
Here $h(t)$ contains the electronic kinetic energy and static or
time-dependent external potentials.
The electron-electron interaction tensor $v^{\rm phys}(t)$ is constructed
from the real, symmetric Coulomb interaction
$v_{\rm C}(\mathbf r,\mathbf r')=|\mathbf r-\mathbf r'|^{-1}$,
with a possible time dependence introduced through adiabatic switching.
The symmetric matrix $Q^{\mathsf T}(t)=Q(t)$ describes the free
phoson dynamics and possible couplings between phoson modes,
including photon-phonon interactions~\cite{Latini2021Phonoritons}.
The vector $l(t)$ represents coherent driving of the phoson
degrees of freedom, while $g^{\rm phys}(t)$ describes the linear
electron-phoson coupling, which may also be switched on adiabatically.

For a specific material and cavity geometry, the microscopic
determination of these parameters, as well as the Lindblad
jump operators introduced below, can be a challenging task.
For the photonic degrees of freedom, one possible approach
is based on quasinormal modes (QNMs), which describe the
resonances of open cavities through complex eigenfrequencies
and spatial field profiles
\cite{Lalanne2018,Kristensen2020}.
Quantized QNM approaches can further provide microscopic
descriptions of light-matter coupling and cavity dissipation,
including dissipative coupling between different modes
\cite{Franke2019,Franke2022}.
Such calculations depend on the specific physical system and
are beyond the scope of the present work.
Here, we take the Hamiltonian parameters and the Lindblad
jump operators as given and focus on developing a general
formalism to describe the dynamics and correlations of
interacting open electron-phoson systems.

The electron-number operator is
\begin{align}
    \hat N_{\rm el}
    &\equiv
    \int d\mathbf x\,
    \hat\psi^\dagger(\mathbf x)\hat\psi(\mathbf x)
    =\sum_n\hat{\bar d}_n^\dagger\hat d_n,
\end{align}
which commutes with the Hamiltonian,
$[\hat N_{\rm el},\hat H]  =0$.

We now specify the Hermitian-adjoint transformations of the
coefficients appearing in Eq.~\eqref{eq:model-H-main}.
With $h=(h_{mn})$, $l=(l_\mu)$, and
$g^{\rm phys}_\mu=(g^{\rm phys}_{\mu,mn})$, the adjoint transformations are
\begin{subequations}
\label{eq:coefficient-adjoints-main}
\begin{align}
    h^{\rm ad}
    &=\mathcal M^{-1}h^\dagger\mathcal M,
    \label{eq:electron-adjoint-main}\\
    v^{\rm phys,ad}_{ijmn}
    &=\sum_{abcd}
    (\mathcal M^{-1})_{ic}
    (\mathcal M^{-1})_{jd}
    v_{abcd}^{\rm phys *}\,\mathcal M_{am}\mathcal M_{bn},
    \label{eq:two-electron-adjoint-main}\\
    Q^{\rm ad}
    &=\kappa^{\mathsf T}Q^*\kappa,
    \\
    l^{\rm ad}
    &=\kappa^{\mathsf T}l^*,
    \\
    g_\mu^{\rm phys,ad}
    &=\sum_{\mu'}\kappa_{{\mu'}\mu}
    \mathcal M^{-1}g_{\mu'}^{\rm phys \dagger}\mathcal M.
\end{align}
\end{subequations}
Consequently, the self-adjointness of the Hamiltonian in Eq.~\eqref{eq:model-H-main} requires
\begin{align}
    &h^{\rm ad}=h,\quad
    v^{\rm phys,ad}=v^{\rm phys},\quad
    Q^{\rm ad}=Q,\nonumber\\
    &l^{\rm ad}=l,\quad
    g^{\rm phys,ad}=g^{\rm phys}.
    \label{eq:model-hermiticity-main}
\end{align}

\subsection{Lindbladian dynamics}

The reduced many-body density matrix $\hat\rho(t)$ of the open
electron-phoson system evolves according to the Lindblad master
equation~\cite{Davies1974Markov,Gorini1976GKSL,Lindblad1976Generator},
\begin{align}
    \frac{d}{dt}\hat\rho(t)
    &=-i[\hat H(t),\hat\rho(t)]
    +\sum_{n=1}^{3}\mathcal D_t^{(n)}[\hat\rho(t)],
    \label{eq:lindblad-master-main}
\end{align}
where the coherent evolution is governed by the many-body
Hamiltonian $\hat H(t)$ in Eq.~\eqref{eq:model-H-main}.
The coupling to the environment is described by three classes
of dissipator superoperators,
\begin{align}
    \mathcal D_t^{(n)}[\hat\rho]
    &=\sum_{\lambda_n}\Big(
    2\hat L_{\lambda_n}^{(n)}(t)
    \hat\rho\hat L_{\lambda_n}^{(n)\dagger}(t)
    \nonumber\\
    &\quad-
    \big\{
    \hat L_{\lambda_n}^{(n)\dagger}(t)
    \hat L_{\lambda_n}^{(n)}(t),
    \hat\rho
    \big\}\Big),
    \label{dissipator}
\end{align}
where $\lambda_n$ labels the dissipative channels within
each class. We consider the following general forms of
the jump operators:
\begin{subequations}
\label{genlind}
\begin{align}
    \hat L_{\lambda_1}^{(1)}(t)
    &=\sum_n \eta_n^{\lambda_1}(t)\hat d_n,
    \label{loss}\\
    \hat L_{\lambda_2}^{(2)}(t)
    &=\sum_n \bar\eta_n^{\lambda_2 *}(t)
    \hat{\bar d}_n^\dagger,
    \label{gain}\\
    \hat L_{\lambda_3}^{(3)}(t)
    &=\sum_{mn}\xi_{mn}^{\lambda_3}(t)
    \hat{\bar d}_m^\dagger\hat d_n
    +\sum_\mu\zeta_\mu^{\lambda_3}(t)
    \hat\phi_\mu.
    \label{phloss}
\end{align}
\end{subequations}
Here $\eta$, $\bar\eta$, $\xi$, and $\zeta$ are complex amplitudes
characterizing the corresponding dissipative channels.

The first two classes describe single-electron loss and gain,
respectively, and can model electron extraction from and injection
into the system through external electrodes or electronic
reservoirs. Such processes are commonly employed in quantum
transport, where the injection and extraction rates are
controlled by the reservoir occupations and their coupling
to the system \cite{Dzhioev2011,Schwarz2016,Jin2020}.

The third class encompasses a broader range of dissipative
processes. When only the fermionic bilinear terms are present,
the jump operators describe particle-number-conserving
electronic transitions, including intraband scattering,
interband electron-hole recombination, and dephasing
\cite{Rosati2014}.
When only the bosonic terms are retained, they describe
photon or phonon loss and gain, depending on the chosen
combinations of bosonic field operators
\cite{Carusotto2013}.
More generally, the simultaneous presence of fermionic
and bosonic contributions within the same jump operator
allows for correlated electron-phoson dissipation.
An important example is the decay of exciton-polaritons,
where the dissipative channel acts on a coherent
superposition of excitonic and photonic components
rather than on the individual constituents separately \cite{Rahmani2026Polaritons,Colas2015,Rahmani2016}.

In the nonorthogonal electronic basis, the Hermitian
conjugates of the first two jump operators involve
the overlap matrix $\mathcal M$. Explicitly,
\begin{align}
    \hat L_{\lambda_1}^{(1)\dagger}(t)
    &=\sum_m
    \left[
    \mathcal M^{-1}\eta^{\lambda_1*}(t)
    \right]_m
    \hat{\bar d}_m^\dagger,
    \\
    \hat L_{\lambda_2}^{(2)\dagger}(t)
    &=\sum_n
    \left[
    \mathcal M^{\mathsf T}\bar\eta^{\lambda_2}(t)
    \right]_n
    \hat d_n,
\end{align}
where $\eta^{\lambda_1}$ and $\bar\eta^{\lambda_2}$
denote the corresponding coefficient vectors.
Likewise, the adjoint of the third jump operator is
obtained from the coefficient transformations
\begin{align}
    (\xi^{\lambda_3})^{\rm ad}
    &=\mathcal M^{-1}
    (\xi^{\lambda_3})^\dagger\mathcal M,
    \\
    (\zeta^{\lambda_3})^{\rm ad}
    &=\kappa^{\mathsf T}(\zeta^{\lambda_3})^*,
\end{align}
as defined for the Hamiltonian coefficients.

Finally, we note the scalar-shift freedom of the
Lindblad jump operators. Under the transformation
\begin{align}\label{gauge1}
    \hat L_{\lambda_3}^{(3)}(t)
    \longrightarrow
    \hat L_{\lambda_3}^{(3)}(t)
    +c_{\lambda_3}(t),
\end{align}
with the Hamiltonian unchanged, the resulting master
equation is equivalent to one with the original jump
operators and a modified Hamiltonian
$\hat H(t)\rightarrow\hat H(t)+\delta\hat H(t)$, where
\begin{align}
    \delta\hat H(t)
    =i\sum_{\lambda_3}
    \Big[
    c_{\lambda_3}^*(t)
    \hat L_{\lambda_3}^{(3)}(t)-
    c_{\lambda_3}(t)
    \hat L_{\lambda_3}^{(3)\dagger}(t)
    \Big].
    \label{eq:lindblad-scalar-shift-main}
\end{align}
Equivalently, the simultaneous transformations
$\hat L_{\lambda_3}^{(3)}
\rightarrow\hat L_{\lambda_3}^{(3)}+c_{\lambda_3}$
and
$\hat H\rightarrow\hat H-\delta\hat H$
leave the Lindblad dynamics invariant.
Since $\delta\hat H$ contains only electronic one-body
and linear phoson operators, this freedom amounts to
a redefinition of the coefficients $h(t)$ and $l(t)$
in Eq.~\eqref{eq:model-H-main}.

\section{Dissipative Hamiltonian on the Keldysh contour}\label{Dissipative Hamiltonian on the Keldysh contour}
We consider an initial density matrix of the form
$\hat\rho(t_0)=Z^{-1}e^{-\beta\hat H^{\rm M}}$,
where $\beta>0$ is the inverse temperature,
$Z=\Tr e^{-\beta\hat H^{\rm M}}$ is the partition function,
and $\hat H^{\rm M}$ is a self-adjoint Matsubara Hamiltonian
with the same electron-phoson operator structure as
Eq.~\eqref{eq:model-H-main}.
The choice $\hat H^{\rm M}=\hat H(t_0)-\mu\hat N_{\rm el}$,
with chemical potential $\mu$, corresponds to a
grand-canonical equilibrium initial state, whereas more
general choices describe quasi-equilibrium initial
states~\cite{Perfetto2019ExcitonicInitial,StefanucciVanLeeuwen2025}.

The initial state and subsequent real-time evolution can be
represented on the L-shaped Keldysh contour
$C=C_-\cup C_+\cup C_{\rm M}$, consisting of the forward
branch $C_-=(t_0,\infty)$, the backward branch
$C_+=(\infty,t_0)$, and the Matsubara branch
$C_{\rm M}=(t_0,t_0-i\beta)$~\cite{Keldysh1965,Danielewicz1984,StefanucciVanLeeuwen2025}.

Time-dependent expectation values obtained from
Eq.~\eqref{eq:lindblad-master-main} can equivalently be
expressed as contour-ordered averages~\cite{stefanucci2024kadanoff},
\begin{align}
    \langle\hat O(t)\rangle
    &\equiv
    \Tr[\hat\rho(t)\hat O(t)]
    \nonumber\\
    &=
    \frac{
    \Tr\left[
    \mathcal T\left\{
    e^{-i\int_C d\bar z\,\hat{\mathcal H}(\bar z)}
    \hat O(t_\pm)
    \right\}
    \right]
    }{
    \Tr\left[
    \mathcal T\left\{
    e^{-i\int_C d\bar z\,\hat{\mathcal H}(\bar z)}
    \right\}
    \right]
    },
    \label{eq:contour-average-main}
\end{align}
where $\mathcal T$ denotes contour ordering.
The dissipative contour Hamiltonian is given by
\begin{align}
    \hat{\mathcal H}(z)
    &=\hat H_{\rm o}(z)
    +2i\theta_-(z)
    \sum_{n=1}^3\sum_{\lambda_n}
    \hat L_{\lambda_n}^{(n)\dagger}(z^*)
    \hat L_{\lambda_n}^{(n)}(z),
    \label{eq:KL-generator-main}
\end{align}
with
\begin{align}
    \hat H_{\rm o}(z)
    &=\hat H(z)
    -is(z)\sum_{n=1}^3\sum_{\lambda_n}
    \hat L_{\lambda_n}^{(n)\dagger}(z)
    \hat L_{\lambda_n}^{(n)}(z).
    \label{eq:Ho-main}
\end{align}

We denote a contour time by $z=t_\pm$ on $C_\pm$
and by $z=t_0-i\tau$ on $C_{\rm M}$.
The branch indicator functions are defined by
$\theta_\pm(z)=1$ for $z\in C_\pm$ and zero otherwise,
and $s(z)=\theta_-(z)-\theta_+(z)$.
For a real-time point $z=t_\pm$, we denote the
corresponding point on the opposite branch by
$z^*=t_\mp$.
The Hamiltonian and jump operators take the same
physical-time values on both real-time branches,
$\hat H(t_\pm)=\hat H(t)$ and
$\hat L_{\lambda_n}^{(n)}(t_\pm)
=\hat L_{\lambda_n}^{(n)}(t)$.
On the Matsubara branch, the jump operators vanish
and $\hat{\mathcal H}(t_0-i\tau)=\hat H^{\rm M}$.

Accordingly, the contour extensions of the Hamiltonian
coefficients $a=h,v,Q,l,g$ are defined by
$a(t_\pm)=a(t)$ and $a(t_0-i\tau)=a^{\rm M}$,
whereas the jump coefficients
$b=\eta,\bar\eta,\xi,\zeta$ satisfy
$b(t_\pm)=b(t)$ and $b(t_0-i\tau)=0$.

We define the contour delta function by $\int_C d\bar z\,\delta(z,\bar z)f(\bar z)=f(z)$.
For arguments on the Matsubara branch,
$\delta(z,z')=i\delta(\tau-\tau')$.
On the real-time branches, their opposite orientations imply
$\delta(z,z'^*)=-\delta(z^*,z')$.
Expressions involving $z^*$ are understood to apply only
to real-time branches, since their dissipative prefactors
vanish on $C_{\rm M}$.

To develop the diagrammatic formulation, we first reorganize
the dissipative contour Hamiltonian into quadratic and
interaction contributions. In particular, the linear phoson
source can be absorbed into a field displacement, allowing
us to express the Hamiltonian in terms of shifted phoson
operators. To construct this displacement consistently
with the dissipative dynamics, we begin by examining
the equation of motion for the phoson one-point function,
\begin{align}
    \Phi_\mu(z)
    \equiv
    \langle\hat\phi_\mu(z)\rangle.
    \label{eq:Phi-main}
\end{align}

Its equation of motion takes the form
\begin{align}
    \left[
        i\alpha^{-1}\frac{d}{dz}
        -\mathcal K(z)
    \right]\Phi(z)
    =
    l(z)+F^{\rm ph}(z).
    \label{eq:Phi-EOM-main}
\end{align}
Here $\mathcal K$ is the effective phoson kernel,
\begin{align}
    \mathcal K(z)
    &\equiv
    Q(z)
    +i\left[
        \Lambda^<(z)-\Lambda^{>}(z)
    \right],
    \label{eq:K-main}
\end{align}
where the purely phosonic dissipative
contributions are characterized by
\begin{align}
    \Lambda^<_{\mu_1\mu_2}(z)
    &\equiv
    \sum_\lambda
    \zeta^\lambda_{\mu_1}(z)
    \zeta^{\lambda,\rm ad}_{\mu_2}(z),
    \label{eq:Lambda-main}
\end{align}
with
$\Lambda^>(z)=\Lambda^{<,\mathsf T}(z)$.

The electron-dependent force in
Eq.~\eqref{eq:Phi-EOM-main} is given by
\begin{align}
    F^{\rm ph}_\mu(z)
    &\equiv
    \sum_{mn}
    \mathcal G^{\rm ph}_{\mu,mn}(z)
    \rho_{nm}(z),
    \label{eq:phoson-force-main}
\end{align}
where
\begin{align}\label{defrho}
    \rho_{mn}(z)
    &\equiv
    \left\langle
        \hat{\bar d}_n^\dagger(z)
        \hat d_m(z)
    \right\rangle
\end{align}
is the one-electron density matrix.
In the biorthogonal representation, it satisfies
$\rho^{\rm ad}=\rho$, or equivalently
$\rho^\dagger\mathcal M=\mathcal M\rho$.

The effective electron-phoson coupling entering
the phoson equation of motion is defined as
\begin{align}
    \mathcal G^{\rm ph}(z)
    &\equiv
    g^{\rm phys}(z)
    +i\left[
        g^{\rm diss}(z)-g^{\rm diss,ad}(z)
    \right],
    \label{eq:effective-gph-main}
\end{align}
where the mixed dissipative contributions are
\begin{align}
    g^{\rm diss}_{\mu,mn}(z)
    &=
    \sum_\lambda
    \zeta^\lambda_\mu(z)
    \xi^{\lambda,\rm ad}_{mn}(z),
    \\
    g^{\rm diss,ad}_{\mu,mn}(z)
    &=
    \sum_\lambda
    \zeta^{\lambda,\rm ad}_\mu(z)
    \xi^\lambda_{mn}(z).
\end{align}

To define the reference phoson propagator on the Keldysh
contour, we must retain the dissipative couplings between
different contour branches, which are not explicit in the
single-time equation of motion above. We therefore
introduce the quadratic contour kernel
\begin{align}
    \mathcal Q(z,z')
    &=\delta(z,z')
    \left\{
        Q(z)
        -is(z)
        [\Lambda^<(z)+\Lambda^>(z)]
    \right\}
    \nonumber\\
    &\quad
    -2i\delta(z,z'^*)
    \left[
        \theta_-(z)\Lambda^<(z)
        -\theta_+(z)\Lambda^>(z)
    \right].
    \label{eq:Q-contour}
\end{align}
This kernel is symmetric under simultaneous
exchange of its phoson indices and contour times, $\mathcal Q_{\mu\mu'}(z,z')
    =\mathcal Q_{\mu'\mu}(z',z)$.
In terms of $\mathcal Q$, Eq.~\eqref{eq:Phi-EOM-main}
can equivalently be expressed as
\begin{align}
    i\alpha^{-1}\frac{d}{dz}\Phi(z)
    -\int_C d\bar z\,
    \mathcal Q(z,\bar z)\Phi(\bar z)
    =l(z)+F^{\rm ph}(z).
    \label{eq:Phi-EOM1-main}
\end{align}

To absorb the prescribed linear phoson source into
a contour displacement, we introduce an auxiliary
Green's function $D^0$ defined by
\begin{align}
    i\alpha^{-1}\frac{d}{dz}D^0(z,z')
    &-\int_C d\bar z\,
    \mathcal Q(z,\bar z)D^0(\bar z,z')
    =\delta(z,z'),
    \label{eq:D0-inverse-main}
\end{align}
subject to the phosonic
Kubo-Martin-Schwinger (KMS) boundary conditions.

The contour displacement is defined as
\begin{align}
    \Delta_\mu(z)
    &\equiv
    \sum_{\mu_1}\int_C d\bar z\,
    D^0_{\mu\mu_1}(z,\bar z)
    l_{\mu_1}(\bar z).
    \label{eq:Delta-contour-main}
\end{align}
On the Matsubara branch, the displacement is
time independent and satisfies
$Q^{\rm M}\Delta^{\rm M}=-l^{\rm M}$.

We then define the shifted phoson field and
its expectation value by
\begin{align}
    \hat\varphi_\mu(z)
    &\equiv
    \hat\phi_\mu(z)-\Delta_\mu(z),
    .
    \label{eq:shifted-field-main}
\end{align}

Using Eq.~\eqref{eq:D0-inverse-main}, the equation
of motion for the shifted one-point function reduces to
\begin{align}
    \left[
        i\alpha^{-1}\frac{d}{dz}
        -\mathcal K(z)
    \right]\langle\hat\varphi(z)\rangle
    =F^{\rm ph}(z).
    \label{eq:shifted-Phi-EOM-main}
\end{align}
Its solution can be expressed in terms of
the reference phoson Green's function $D^0$ as
\begin{align}
    \langle\hat\varphi_\mu(z)\rangle
    &=\sum_{\mu'mn}
    \int_C d\bar z\,
    D^0_{\mu\mu'}(z,\bar z)
    \mathcal G^{\rm ph}_{\mu',mn}(\bar z)
    \rho_{nm}(\bar z).
    \label{eq:shifted-Phi-solution-main}
\end{align}

Having absorbed the prescribed linear phoson source
into the displacement, while leaving the electron-induced
force explicit, we now return to the dissipative contour
Hamiltonian in Eq.~\eqref{eq:KL-generator-main}
and express it in terms of the shifted field
$\hat\varphi$.
Exploiting the scalar-shift freedom of the Lindblad
jump operators discussed in Eq.~\eqref{eq:lindblad-scalar-shift-main}, we reorganize the
contour Hamiltonian into quadratic and interaction
contributions. Normal ordering the electronic operators
and symmetrizing the quadratic phoson terms yield an
equivalent representation within the contour-ordered
exponential. Up to scalar contributions that cancel
in normalized contour averages, we obtain
\begin{align}
    \int_C dz\,\hat{\mathcal H}(z)
    &\doteq
    \int_C dz\,
    \Big[
        \hat{\mathcal H}_{0,\rm el}(z)
        +\hat{\mathcal H}_{0,\rm ph}(z)
        \nonumber\\
    &\hspace{3.7em}
        +\hat{\mathcal H}_{{\rm el}\text{-}{\rm el}}(z)
        +\hat{\mathcal H}_{{\rm el}\text{-}{\rm ph}}(z)
    \Big],
    \label{eq:KL-vertex-decomposition-main}
\end{align}
where $\doteq$ denotes equivalence within
contour-ordered expressions.

The quadratic electronic contribution reads
\begin{align}
    \hat{\mathcal H}_{0,\rm el}(z)
    &=
    \sum_{mn}\int_C dz'\,
    \mathfrak h_{mn}(z',z)
    \hat{\bar d}_m^\dagger(z'^+)\hat d_n(z),
    \label{eq:quadratic-electron-generator-main}
\end{align}
with
\begin{align}
    \mathfrak h_{mn}(z',z)
    &=\delta(z',z)h^{(\Delta)}_{mn}(z)
    \nonumber\\
    &\quad
    -is(z)\delta(z',z)
    \Big[
        V_{mn}^{\rm loss}(z)
        -V_{mn}^{\rm gain}(z)
        +V_{mn}^{\rm diss}(z)
    \Big]
    \nonumber\\
    &\quad
    +2i\delta(z',z^*)
    \Big[
        \theta_-(z)V_{mn}^{\rm loss}(z)
        +\theta_+(z)V_{mn}^{\rm gain}(z)
    \Big].
    \label{eq:hfrak-contour-main}
\end{align}
We denote by $z^+$ a point infinitesimally later than $z$ along
the contour.

The displaced one-electron Hamiltonian is defined by
\begin{align}
    h^{(\Delta)}_{mn}(z)
    &=
    h_{mn}(z)
    +\sum_\mu
    \Delta_\mu(z)
    \mathcal G^{\rm el}_{\mu,mn}(z),
    \label{eq:shifted-h-main}
\end{align}
where the effective electron-phoson coupling entering
the electronic equation of motion is
\begin{align}
    \mathcal G^{\rm el}(z)
    &\equiv
    g^{\rm phys}(z)
    -i\left[
        g^{\rm diss}(z)-g^{\rm diss,ad}(z)
    \right].
    \label{eq:effective-gp-main}
\end{align}

In addition to the displaced one-electron Hamiltonian
$h^{(\Delta)}$, the 
$\mathfrak h$ contains dissipative potentials
arising from single-electron loss and gain,
\begin{subequations}\label{vlossgain}
\begin{align}
    V_{mn}^{\rm loss}(z)
    &=\sum_{\lambda_1}
    \left[
    \mathcal M^{-1}\eta^{\lambda_1*}(z)
    \right]_m
    \eta_n^{\lambda_1}(z),
    \label{Vloss}\\
    V_{mn}^{\rm gain}(z)
    &=\sum_{\lambda_2}
    \bar\eta_m^{\lambda_2*}(z)
    \left[
    \mathcal M^{\mathsf T}\bar\eta^{\lambda_2}(z)
    \right]_n.
    \label{Vgain}
\end{align}
\end{subequations}
as well as the contribution from the third jump operators,
\begin{align}
    V_{mn}^{\rm diss}(z)
    &=\frac12\sum_j
    v_{mjnj}^{\rm diss}(z),
    \label{cpvdiss1}
\end{align}
with
\begin{align}
    v^{\rm diss}_{ijmn}(z)
    &\equiv
    2\sum_\lambda
    \xi^{\lambda,\rm ad}_{in}(z)
    \xi^\lambda_{jm}(z).
    \label{eq:v-diss-main}
\end{align}

The quadratic and linear phosonic contributions are
\begin{align}
    \hat{\mathcal H}_{0,\rm ph}(z)
    &=
    \frac12\sum_{\mu\mu'}
    \int_C dz'\,
    \mathcal Q_{\mu\mu'}(z,z')
    \hat\varphi_\mu(z)
    \hat\varphi_{\mu'}(z')
    \nonumber\\
    &\quad
    +\sum_\mu
    l_\mu^{(\Delta)}(z)\hat\varphi_\mu(z),
    \label{eq:quadratic-phoson-generator-main}
\end{align}
where the displaced linear coefficient satisfies
\begin{align}
    l^{(\Delta)}(z)
    &=l(z)+\mathcal K(z)\Delta(z)=i\alpha^{-1}
    \frac{d}{dz}\Delta(z).
    \label{eq:shift-linear-main}
\end{align}

The electron-electron interaction contribution
takes the form
\begin{align}
    \hat{\mathcal H}_{{\rm el}\text{-}{\rm el}}(z)
    &=
    \frac12\sum_{ijmn}\int_C dz'\,
    v_{ijmn}(z,z')
    \nonumber\\
    &\quad\times
    \hat{\bar d}_i^\dagger(z'^+)
    \hat{\bar d}_j^\dagger(z^+)
    \hat d_m(z)\hat d_n(z').
    \label{eq:v-vertex}
\end{align}
The resulting interaction tensor obeys $v_{ijmn}(z,z')=v_{jinm}(z',z)$ and is given by
\begin{align}
    v_{ijmn}(z,z')
    &=\delta(z,z')v^{\rm phys}_{ijmn}(z)
    \nonumber\\
    &\quad
    -is(z)
    \left[
        \delta(z,z')
        +\delta(z,z'^*)
    \right]
    v^{\rm s,diss}_{ijmn}(z)
    \nonumber\\
    &\quad
    -i\delta(z,z'^*)
    v^{\rm a,diss}_{ijmn}(z),
    \label{eq:v-contour}
\end{align}
where
\begin{align}
    v^{\rm s/a,diss}_{ijmn}(z)
    &\equiv
    \frac12
    \left[
        v^{\rm diss}_{ijmn}(z)
        \pm v^{\rm diss}_{jinm}(z)
    \right].
    \label{eq:v-diss-symmetry-main}
\end{align}

Similarly, the electron-phoson interaction contribution
can be written as
\begin{align}
    \hat{\mathcal H}_{{\rm el}\text{-}{\rm ph}}(z)
    &=
    \sum_{\mu mn}\int_C dz'\,
    g_{\mu,mn}(z,z')
    \hat\varphi_\mu(z)
    \hat{\bar d}_m^\dagger(z'^+)\hat d_n(z'),
    \label{eq:mixed-vertex-main}
\end{align}
where the mixed $g_{\mu,mn}(z,z')$ incorporates both
the physical electron-phoson coupling and
the dissipative contributions,
\begin{align}
    &{}\quad g_{\mu,mn}(z,z')
    \nonumber\\
    &=
    \delta(z,z')
        g^{\rm phys}_{\mu,mn}(z)
    \nonumber\\
    &\quad -is(z)
    \delta(z,z')
        \left[
            g^{\rm diss}_{\mu,mn}(z)
            +g^{\rm diss,ad}_{\mu,mn}(z)
        \right]
    \nonumber\\
    &\quad
    -2i\delta(z,z'^*)
    \left[
        \theta_-(z)g^{\rm diss}_{\mu,mn}(z)
        -\theta_+(z)g^{\rm diss,ad}_{\mu,mn}(z)
    \right].
    \label{eq:g-contour}
\end{align}
In Eqs.~\eqref{eq:v-contour} and
\eqref{eq:g-contour}, the coefficients
$v^{\rm phys}_{ijmn}(z)$ and $g^{\rm phys}_{\mu,mn}(z)$ denote the
physical couplings introduced in
Eq.~\eqref{eq:model-H-main}.

Integrating the mixed kernel over its electronic
or phosonic time argument yields, respectively,
\begin{align}
    \int_C dz'\,g_{\mu,mn}(z,z')
    &=\mathcal G^{\rm ph}_{\mu,mn}(z),
    \\
    \int_C dz'\,g_{\mu,mn}(z',z)
    &=\mathcal G^{\rm el}_{\mu,mn}(z).
    \label{eq:g-endpoints}
\end{align}

\section{Dissipative NEGF Formalism}\label{Dissipative NEGF Formalism}

\begin{widetext}

\subsection{Mixed Green's functions and diagrammatic expansions}

For $n=0,1,\ldots$ and $m=0,\frac12,1,\ldots$, we define
the mixed $(n+m)$-particle nonequilibrium Green's function
(NEGF) as
\begin{align}
 &M_{n,m}\left(
 i_1z_1,\ldots,i_nz_n;
 j_1z'_1,\ldots,j_nz'_n\mid
 \mu_1\bar z_1,\ldots,\mu_{2m}\bar z_{2m}\right)
 \nonumber\\
 &=i^{-(n+m)}
 \left\langle
 \hat d_{i_1}(z_1)\cdots\hat d_{i_n}(z_n)
 \hat{\bar d}_{j_n}^\dagger(z_n'^+)
 \cdots\hat{\bar d}_{j_1}^\dagger(z_1'^+)
 \hat\varphi_{\mu_1}(\bar z_1)\cdots
 \hat\varphi_{\mu_{2m}}(\bar z_{2m})
 \right\rangle.
 \label{eq:M-definition}
\end{align}
Here $M_{0,0}=1$ and $i^{1/2}=e^{i\pi/4}$.
The semicolon separates electronic annihilation and creation
arguments, while the vertical bar separates the electronic
and phosonic arguments.

The Green's function is antisymmetric under the exchange
of any two electronic annihilation arguments or any two
creation arguments, and symmetric under permutations of
the phoson arguments.
These permutation properties allow any argument to be
brought to the first position within its respective sector,
with the appropriate fermionic sign.
In the following, the electronic equations of motion
require $n\geq1$, whereas the phosonic equation requires
$m\geq\frac12$.

Taking the contour-time derivatives of
Eq.~\eqref{eq:M-definition} with respect to $z_1$,
$z'_1$, and $\bar z_1$ yields the
Martin-Schwinger hierarchy (MSH)
~\cite{PhysRev.115.1342,StefanucciVanLeeuwen2025}.
The electron-electron interaction kernel $v(z,z')$
couples $M_{n,m}$ to $M_{n+1,m}$.
The mixed electron-phoson vertex $g(z,z')$ couples
$M_{n,m}$ to $M_{n,m+1/2}$ when acting at an electronic
endpoint, and to $M_{n+1,m-1/2}$ when acting at a
phosonic endpoint.
The hierarchy is supplemented by the corresponding
fermionic and bosonic KMS
boundary conditions.

For the jump operators introduced above, we derive
the following MSH equations in a similar way as in Ref.~\cite{zhang2026nhbse}.
Differentiation with respect to the first electronic
annihilation time gives
\begin{align}
 &i\frac{d}{dz_1}
 M_{n,m}\left(
 i_1z_1,\ldots,i_nz_n;
 j_1z'_1,\ldots,j_nz'_n\mid
 \mu_1\bar z_1,\ldots,\mu_{2m}\bar z_{2m}\right)
 \nonumber\\
 &\quad-\sum_k\int_{C}dz\,
 \mathfrak h_{i_1k}(z_1,z)
 M_{n,m}\left(
 kz,i_2z_2,\ldots,i_nz_n;
 j_1z'_1,\ldots,j_nz'_n\mid
 \mu_1\bar z_1,\ldots,\mu_{2m}\bar z_{2m}\right)
 \nonumber\\
 &=\sum_{l=1}^n{(-1)}^{1+l}
 \delta_{i_1j_l}\delta(z_1,z'_l)
 M_{n-1,m}(
 \stackrel{\sqcap}{i_1z_1},i_2z_2,\ldots,i_nz_n;
 j_1z'_1,\ldots,\stackrel{\sqcap}{j_lz'_l},\ldots,j_nz'_n\mid
 \mu_1\bar z_1,\ldots,\mu_{2m}\bar z_{2m})
 \nonumber\\
 &\quad-i \sum_{jkl}
 \int_{C}dz\,v_{i_1jkl}(z,z_1)
 M_{n+1,m}\left(
 lz_1,i_2z_2,\ldots,i_nz_n,kz;
 j_1z'_1,\ldots,j_nz'_n,jz\mid
 \mu_1\bar z_1,\ldots,\mu_{2m}\bar z_{2m}\right)
 \nonumber\\
 &\quad+i^{1/2}\sum_{\mu k}
 \int_{C}dz\,g_{\mu,i_1k}(z,z_1)
 M_{n,m+1/2}\left(
 kz_1,i_2z_2,\ldots,i_nz_n;
 j_1z'_1,\ldots,j_nz'_n\mid
 \mu_1\bar z_1,\ldots,\mu_{2m}\bar z_{2m},\mu z\right).
 \label{eq:M-left}
\end{align}
Similarly, differentiation with respect to the first
electronic creation time yields
\begin{align}
 &-i\frac{d}{dz'_1}
 M_{n,m}\left(
 i_1z_1,\ldots,i_nz_n;
 j_1z'_1,\ldots,j_nz'_n\mid
 \mu_1\bar z_1,\ldots,\mu_{2m}\bar z_{2m}\right)
 \nonumber\\
 &\quad-\sum_k\int_{C}dz\,
 M_{n,m}\left(
 i_1z_1,\ldots,i_nz_n;
 kz,j_2z'_2,\ldots,j_nz'_n\mid
 \mu_1\bar z_1,\ldots,\mu_{2m}\bar z_{2m}\right)
 \mathfrak h_{kj_1}(z,z'_1)
 \nonumber\\
 &=\sum_{l=1}^n{(-1)}^{1+l}
 \delta_{i_lj_1}\delta(z_l,z'_1)
 M_{n-1,m}(
 i_1z_1,\ldots,\stackrel{\sqcap}{i_lz_l},\ldots,i_nz_n;
 \stackrel{\sqcap}{j_1z'_1},j_2z'_2,\ldots,j_nz'_n\mid
 \mu_1\bar z_1,\ldots,\mu_{2m}\bar z_{2m})
 \nonumber\\
 &\quad-i \sum_{abc}
 \int_{C}dz\,v_{abcj_1}(z,z'_1)
 M_{n+1,m}\left(
 i_1z_1,\ldots,i_nz_n,cz;
 az'_1,j_2z'_2,\ldots,j_nz'_n,bz\mid
 \mu_1\bar z_1,\ldots,\mu_{2m}\bar z_{2m}\right)
 \nonumber\\
 &\quad+i^{1/2}\sum_{\mu k}
 \int_{C}dz\,g_{\mu,kj_1}(z,z'_1)
 M_{n,m+1/2}\left(
 i_1z_1,\ldots,i_nz_n;
 kz'_1,j_2z'_2,\ldots,j_nz'_n\mid
 \mu_1\bar z_1,\ldots,\mu_{2m}\bar z_{2m},\mu z\right).
 \label{eq:M-right}
\end{align}
Finally, the equation of motion with respect to
the first phoson time reads
\begin{align}
 &\sum_\mu i(\alpha^{-1})_{\mu_1\mu}
 \frac{d}{d\bar z_1}
 M_{n,m}\left(
 i_1z_1,\ldots,i_nz_n;
 j_1z'_1,\ldots,j_nz'_n\mid
 \mu\bar z_1,\mu_2\bar z_2,\ldots,\mu_{2m}\bar z_{2m}\right)
 \nonumber\\
 &\quad-\sum_\mu\int_{C}dz\,
 \mathcal Q_{\mu_1\mu}(\bar z_1,z)
 M_{n,m}\left(
 i_1z_1,\ldots,i_nz_n;
 j_1z'_1,\ldots,j_nz'_n\mid
 \mu z,\mu_2\bar z_2,\ldots,\mu_{2m}\bar z_{2m}\right)
 \nonumber\\
 &=\sum_{l=2}^{2m}
 \delta_{\mu_1\mu_l}\delta(\bar z_1,\bar z_l)
 M_{n,m-1}(
 i_1z_1,\ldots,i_nz_n;
 j_1z'_1,\ldots,j_nz'_n\mid
 \stackrel{\sqcap}{\mu_1\bar z_1},
 \mu_2\bar z_2,\ldots,\stackrel{\sqcap}{\mu_l\bar z_l},
 \ldots,\mu_{2m}\bar z_{2m})
 \nonumber\\
 &\quad-i^{1/2}\sum_{ab}
 \int_{C}dz\,g_{\mu_1,ab}(\bar z_1,z)
 M_{n+1,m-1/2}(
 i_1z_1,\ldots,i_nz_n,bz;
 j_1z'_1,\ldots,j_nz'_n,az\mid
 \stackrel{\sqcap}{\mu_1\bar z_1},
 \mu_2\bar z_2,\ldots,\mu_{2m}\bar z_{2m}).
 \label{eq:M-ph}
\end{align}
In these equations, the symbol $\sqcap$ above an
argument indicates that the corresponding argument
is omitted.

When the full interaction kernels $v(z,z')$ and
$g(z,z')$ vanish, the MSH reduces to a noninteracting
hierarchy involving only lower-order Green's functions.
Its solution $M^0_{n,m}$ is given by Wick's
theorem~\cite{stefanucci2024kadanoff,zhang2026nhbse},
which factorizes the mixed Green's function into
fermionic and phosonic contractions:
\begin{align}
 M^0_{n,m}
 &(i_1z_1,\ldots,i_nz_n;
 j_1z'_1,\ldots,j_nz'_n\mid
 \mu_1\bar z_1,\ldots,\mu_{2m}\bar z_{2m})
 \nonumber\\
 &=\left[
 \sum_{\pi\in S_n}(-1)^{N(\pi)}
 \prod_{a=1}^nG^0_{i_a j_{\pi(a)}}(z_a,z'_{\pi(a)})
 \right]
 \left[
 \sum_{\mathcal P_{2m}}
 \prod_{(b,c)\in\mathcal P_{2m}}
 D^0_{\mu_b\mu_c}(\bar z_b,\bar z_c)
 \right].
 \label{eq:mixed-wick}
\end{align}
Here $S_n$ is the permutation group of $n$ electronic
indices, $N(\pi)$ denotes the parity of the permutation
$\pi$, and $\mathcal P_{2m}$ runs over all distinct
complete pairings of the $2m$ phoson arguments,
each counted once.
For half-integer $m$, the noninteracting Green's
function vanishes, $M^0_{n,m}=0$.

The noninteracting phoson propagator $D^0$ is defined
by Eq.~\eqref{eq:D0-inverse-main}, while the
noninteracting electron propagator $G^0$ satisfies
\begin{align}
 i\frac{d}{dz}G^0(z,z')
 -\int_{C}d\bar z\,
 \mathfrak h(z,\bar z)G^0(\bar z,z')
 =\delta(z,z').
 \label{eq:G0}
\end{align}
The fermionic factor in Eq.~\eqref{eq:mixed-wick}
is a determinant, whereas the phosonic factor
is a sum over pairings, equivalently expressed
as a hafnian~\cite{StefanucciVanLeeuwen2025}.

The MSH derived above, together with the corresponding
Wick factorization, establishes that the open electron-phoson
system admits the same diagrammatic rules as its closed
counterpart~\cite{StefanucciVanLeeuwen2025}.
In particular, the diagrammatic topology, combinatorial
factors, and fermionic sign rules remain unchanged.
Environmental effects are entirely incorporated into the
elementary building blocks, namely the noninteracting
propagators $G^0$ and $D^0$, the electron-electron
interaction kernel $v$, and the electron-phoson coupling
kernel $g$.
Consequently, the diagrammatic many-body techniques
developed for closed electron-boson systems can be
extended to the present open-system framework using
these generalized propagators and interaction vertices.

We define the full electron Green's function $G$ and
the connected phoson Green's function $D$ as
\begin{align}
    G_{ij}(z,z')
    &\equiv M_{1,0}(iz;jz'),\\
    D_{\mu\mu'}(z,z')
    &\equiv M_{0,1}(\mu z,\mu'z')
    -\frac{1}{i}
    \langle\hat\varphi_\mu(z)\rangle\langle\hat\varphi_{\mu'}(z')\rangle.
    \label{eq:M01-connected}
\end{align}
Their diagrammatic expansions can be resummed into
the Dyson equations
\begin{align}
    G(z,z')
    &=G^0(z,z')
    +\int_C d\bar z\,d\bar z'\,
    G^0(z,\bar z)
    \Sigma(\bar z,\bar z')
    G(\bar z',z'),\\
    D(z,z')
    &=D^0(z,z')
    +\int_C d\bar z\,d\bar z'\,
    D^0(z,\bar z)
    \Pi(\bar z,\bar z')
    D(\bar z',z'),
\end{align}
where $\Sigma$ and $\Pi$ denote the irreducible
electron and phoson self-energies, respectively.

\end{widetext}

\section{Phoson Kadanoff-Baym equations}
\label{Kadanoff-Baym equations}
\label{sec:phoson-KBE}

The Wick expansion of the mixed Green's functions
provides the Feynman rules for constructing the
irreducible electron and phoson self-energies,
$\Sigma[G^0,D^0,v,g]$ and $\Pi[G^0,D^0,v,g]$.
As in the closed-system formulation, the
self-energies can be expressed in terms of fully
dressed electron and phoson Green's functions
through skeleton diagrams, namely diagrams
without self-energy insertions~\cite{StefanucciVanLeeuwen2025}.

Since the phoson Green's function is defined
as a connected propagator, the electronic
self-energy contains a separate contribution
from the shifted phoson one-point function.
We therefore write
\begin{align}
    \Sigma
    &=\Sigma^{\rm Eh}+\Sigma^{\rm s},\quad
    \Pi
    =\Pi^{\rm s}.
\end{align}
Here $\Sigma^{\rm Eh}$ is the Ehrenfest
contribution associated with
$\langle\hat\varphi\rangle$
defined in Eq.~\eqref{eq:shifted-field-main}.
The skeleton self-energies
$\Sigma^{\rm s}[G,D,v,g]$ and
$\Pi^{\rm s}[G,D,v,g]$ are functionals
of the fully dressed propagators and
the generalized interaction kernels.
The separation of $\Sigma^{\rm Eh}$
avoids double counting the phoson
one-point contribution.

The corresponding contour equation
for the connected phoson Green's
function reads
\begin{align}
    i\alpha^{-1}\frac{d}{dz}D(z,z')
    &-\int_C d\bar z\,
    [\mathcal Q(z,\bar z)
    +\Pi^{\rm s}(z,\bar z)]
    D(\bar z,z')
    \nonumber\\
    &=\delta(z,z')I.
    \label{eom2}
\end{align}
The KBEs follow by projecting this equation
onto the real-time and Matsubara branches
of the Keldysh contour.
The electronic KBEs have the general structure
established in Ref.~\cite{zhang2026nhbse};
their explicit form for the present
electron-phoson system is given for
completeness in Appendix~\ref{app:electron-KBE}.
The equal-time equations for the electronic
density matrix and phoson covariance are
discussed in Sec.~\ref{sec:equal-time-KBE}.

For a contour two-time function $X$,
we denote its lesser and greater components by
$X^{\lessgtr}(t,t')=X(t_{\mp},t'_{\pm})$.
The mixed components are
$X^{\rceil}(t,\tau)=X(t_\pm,t_0-i\tau)$
and
$X^{\lceil}(\tau,t)=X(t_0-i\tau,t_\pm)$,
while
$X^{\rm M}(\tau,\tau')
=X(t_0-i\tau,t_0-i\tau')$
is the Matsubara component.
These conventions apply to both electronic
and phosonic Green's functions and
self-energies.
We define the real-time and imaginary-time
convolutions as
\begin{align}
    [A\cdot B](t,t')
    &\equiv
    \int_{t_0}^{\infty}d\bar t\,
    A(t,\bar t)B(\bar t,t'),
    \\
    [A\star B](t,t')
    &\equiv
    -i\int_0^\beta d\bar\tau\,
    A(t,\bar\tau)B(\bar\tau,t'),
\end{align}
where the arguments are understood to
belong to the appropriate contour branches.

Applying the Langreth rules
~\cite{Langreth1976,StefanucciVanLeeuwen2025}
to Eq.~\eqref{eom2}, we obtain the
phoson KBEs.
The purely phosonic dissipative
contributions are already incorporated
into the quadratic contour kernel
$\mathcal Q$ and the effective
single-time matrix $\mathcal K$.
For notational simplicity, we henceforth
omit the superscript ${\rm s}$ from
$\Pi^{\rm s}$, with $\Pi$ understood as
the skeleton self-energy evaluated
with fully dressed propagators.

The left- and right-acting lesser
and greater phoson KBEs are
\begin{subequations}
\begin{align}
    &\left[
        i\alpha^{-1}\frac{d}{dt}
        -\mathcal K(t)
    \right]D^\lessgtr(t,t')
    +2i\Lambda^\lessgtr(t)
    D^{\rm A}(t,t')
    \nonumber\\
    &\qquad=
    \big[
        \Pi^{\rm R}\cdot D^\lessgtr
        +\Pi^\lessgtr\cdot D^{\rm A}
        +\Pi^{\rceil}\star D^{\lceil}
    \big](t,t'),
    \label{eq:phoson-KBE-left-main}
    \\
    &D^\lessgtr(t,t')
    \left[
        -i\frac{\overleftarrow d}{dt'}
        \alpha^{-1}
        -\mathcal K^{\mathsf T}(t')
    \right]
    +2iD^{\rm R}(t,t')
    \Lambda^\lessgtr(t')
    \nonumber\\
    &\qquad=
    \big[
        D^\lessgtr\cdot\Pi^{\rm A}
        +D^{\rm R}\cdot\Pi^\lessgtr
        +D^{\rceil}\star\Pi^{\lceil}
    \big](t,t').
    \label{eq:phoson-KBE-right-main}
\end{align}
\end{subequations}

The retarded phoson Green's function satisfies
\begin{align}
    \left[
        i\alpha^{-1}\frac{d}{dt}
        -\mathcal K(t)
    \right]D^{\rm R}(t,t')
    &=\delta(t-t')I
    \nonumber\\
    &\quad+
    [\Pi^{\rm R}\cdot D^{\rm R}](t,t'),
    \label{eq:phoson-R-left-app}
\end{align}
while the advanced component is related
to the retarded one by
\begin{align}
    D^{\rm A}(t,t')
    &=[D^{\rm R}(t',t)]^{\mathsf T}.
\end{align}
This relation follows from the phoson
commutation relations and remains valid
in the general transformed phoson
representation.

The right-mixing component obeys
\begin{subequations}
\begin{align}
    &\left[
        i\alpha^{-1}\frac{d}{dt}
        -\mathcal K(t)
    \right]D^{\rceil}(t,\tau)
    \nonumber\\
    &\qquad=
    \big[
        \Pi^{\rm R}\cdot D^{\rceil}
        +\Pi^{\rceil}\star D^{\rm M}
    \big](t,\tau),
    \label{eq:phoson-rceil-real-app}
    \\
    &D^{\rceil}(t,\tau)
    \left[
        \frac{\overleftarrow d}{d\tau}
        \alpha^{-1}-Q^{\rm M}
    \right]
    \nonumber\\
    &\qquad=
    \big[
        D^{\rm R}\cdot\Pi^{\rceil}
        +D^{\rceil}\star\Pi^{\rm M}
    \big](t,\tau).
    \label{eq:phoson-rceil-M-app}
\end{align}
\end{subequations}

Similarly, the left-mixing component satisfies
\begin{subequations}
\begin{align}
    &\left[
        -\alpha^{-1}\frac{d}{d\tau}
        -Q^{\rm M}
    \right]D^{\lceil}(\tau,t)
    \nonumber\\
    &\qquad=
    \big[
        \Pi^{\rm M}\star D^{\lceil}
        +\Pi^{\lceil}\cdot D^{\rm A}
    \big](\tau,t),
    \label{eq:phoson-lceil-M-app}
    \\
    &D^{\lceil}(\tau,t)
    \left[
        -i\frac{\overleftarrow d}{dt}
        \alpha^{-1}
        -\mathcal K^{\mathsf T}(t)
    \right]
    \nonumber\\
    &\qquad=
    \big[
        D^{\lceil}\cdot\Pi^{\rm A}
        +D^{\rm M}\star\Pi^{\lceil}
    \big](\tau,t).
    \label{eq:phoson-lceil-real-app}
\end{align}
\end{subequations}

Finally, the Matsubara phoson
Green's function satisfies
\begin{subequations}
\begin{align}
    &\left[
        -\alpha^{-1}\frac{d}{d\tau}
        -Q^{\rm M}
    \right]D^{\rm M}(\tau,\tau')
    \nonumber\\
    &\qquad=
    i\delta(\tau-\tau')I
    +[\Pi^{\rm M}\star D^{\rm M}]
    (\tau,\tau'),
    \label{eq:phoson-M-left-app}
    \\
    &D^{\rm M}(\tau,\tau')
    \left[
        \frac{\overleftarrow d}{d\tau'}
        \alpha^{-1}-Q^{\rm M}
    \right]
    \nonumber\\
    &\qquad=
    i\delta(\tau-\tau')I
    +[D^{\rm M}\star\Pi^{\rm M}]
    (\tau,\tau').
    \label{eq:phoson-M-right-app}
\end{align}
\end{subequations}

\section{Equal-time Kadanoff-Baym equations}
\label{sec:equal-time-KBE}

We now turn from two-time propagation to the equations
for the one-electron density matrix and the connected
phoson covariance. These equal-time equations make
explicit the interplay of self-consistent mean fields,
instantaneous dissipative scattering, and the correlation
collision integrals. We first discuss the electronic
mean-field contributions, and then the phoson equation.

\subsection{Electronic density matrix}
\label{subsec:equal-time-electron}

We examine how the electronic mean-field
self-energy renormalizes the generalized one-body
kernel and enters the equal-time dynamics.
As discussed in Sec.~\ref{sec:phoson-KBE},
the electronic self-energy is separated into
the Ehrenfest one-point contribution and the
skeleton self-energy. The latter contains the
Hartree, Fock, and correlation contributions,
while the complete mean-field self-energy
includes the Ehrenfest term,
\begin{subequations}
\label{eq:electron-selfenergy-split-main}
\begin{align}
    \Sigma^{\rm s}
    &=\Sigma^{\rm H}+\Sigma^{\rm F}
    +\Sigma^{\rm c},
    \\
    \Sigma^{\rm MF}
    &=\Sigma^{\rm H}+\Sigma^{\rm F}
    +\Sigma^{\rm Eh}.
\end{align}
\end{subequations}
Thus, the full self-energy satisfies
$\Sigma=\Sigma^{\rm MF}+\Sigma^{\rm c}$.
The Ehrenfest contribution is excluded
from the skeleton self-energy to avoid
double counting.

\begin{figure}
    \centering
    \includegraphics[width=1.0\linewidth]{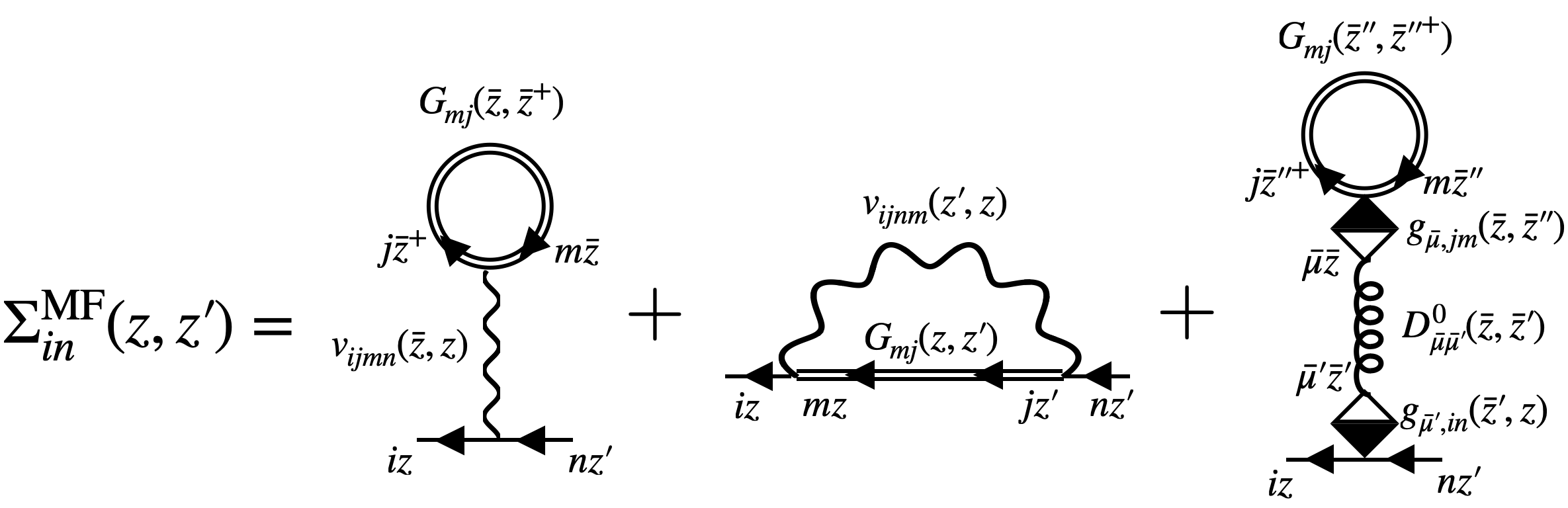}
    \caption{
Electronic mean-field self-energy diagrams:
Hartree, Fock, and Ehrenfest.
Solid double lines denote dressed electronic Green's
functions $G$, wavy lines the generalized electron-electron
interaction kernels $v$, and coiled lines the reference
phoson propagator $D^0$.
Diamonds represent the generalized electron-phoson
vertices $g$.
The mixed vertex $g_{\mu,mn}(z,z')$ has a phoson
endpoint $(\mu,z)$ and an electronic endpoint at $z'$,
with outgoing index $m$ and incoming index $n$.
The interaction vertex $v_{ijmn}(z,z')$ connects
the electronic endpoints $(i,n,z')$ and $(j,m,z)$,
where $i,j$ are outgoing indices and $m,n$ are
incoming indices.
The Hartree and Ehrenfest contributions are local
on the contour, whereas the dissipative Fock
contribution can connect different Keldysh branches.
The Ehrenfest diagram follows from expanding the
shifted phoson one-point function
$\langle\hat\varphi\rangle$ in terms of $D^0$
and the electronic density.
}
    \label{fig:electron-MF-diagrams}
\end{figure}

We further decompose the
mean-field self-energy according to the
origin of its interaction vertices,
\begin{align}
    \Sigma^{\rm MF}
    &=\Sigma^{\rm MF,phys}
    +\Sigma^{\rm MF,diss}.
    \label{eq:MF-phys-diss-split-main}
\end{align}
The physical contributions originate from
the electron-electron and electron-phoson
couplings $v^{\rm phys}$ and $g^{\rm phys}$
in Eq.~\eqref{eq:model-H-main}, whereas
the dissipative contributions arise from
the Lindblad jump operators.
The Hartree-Fock and Ehrenfest diagrams are constructed
using dressed electronic Green's functions
and the generalized interactions
$v(z,z')$ and $g(z,z')$ of Eqs.~\eqref{eq:v-contour} and \eqref{eq:g-contour}, respectively.
Their explicit expressions are given in
Appendix~\ref{app:electron-KBE}.
The three mean-field contributions are illustrated
diagrammatically in Fig.~\ref{fig:electron-MF-diagrams}.

The physical Hartree and Fock self-energies
are local on the contour and combine into
the Hartree-Fock potential
\begin{align}
    V^{\rm HF,phys}_{in}(z)
    &\equiv
    \sum_{jm}
    \left[
        v^{\rm phys}_{ijmn}(z)
        -v^{\rm phys}_{ijnm}(z)
    \right]\rho_{mj}(z).
    \label{eq:HF-potential-main}
\end{align}
The physical Ehrenfest contribution is
also local and gives
\begin{align}
    V^{\rm Eh,phys}_{in}(z)
    &\equiv
    \sum_\mu
    g^{\rm phys}_{\mu,in}(z)
    \langle\hat\varphi_\mu(z)\rangle.
    \label{eq:Eh-phys-potential-main}
\end{align}
Consequently, the physical mean-field
self-energy takes the form
\begin{align}
    \Sigma^{\rm MF,phys}_{in}(z',z)
    &=\delta(z',z)
    \left[
        V^{\rm HF,phys}_{in}(z)
        +V^{\rm Eh,phys}_{in}(z)
    \right].
    \label{eq:MF-phys-local-main}
\end{align}

The dissipative Hartree self-energy is
likewise local on the contour.
Integrating the generalized interaction
kernel over the internal contour time
yields the self-adjoint potential
\begin{align}
    V^{\rm H,diss}_{in}(z)
    &\equiv
    -i\sum_{jm}
    v^{\rm a,diss}_{ijmn}(z)
    \rho_{mj}(z).
    \label{vadiss}
\end{align}
Similarly, the dissipative Ehrenfest
contribution gives
\begin{align}
    V^{\rm Eh,diss}_{in}(z)
    &\equiv
    -i\sum_\mu
    \left[
        g^{\rm diss}_{\mu,in}(z)
        -g^{\rm diss,ad}_{\mu,in}(z)
    \right]
    \langle\hat\varphi_\mu(z)\rangle,
    \label{eq:Eh-diss-potential-main}
\end{align}
both potentials are self-adjoint and have no
explicit dependence on the real-time
branch.

In contrast, the dissipative Fock
self-energy retains the same-branch
and opposite-branch structures of
the generalized interaction kernel.
The dissipative mean-field self-energy
can therefore be expressed as
\begin{align}
    \Sigma^{\rm MF,diss}_{in}(z',z)
    &=\delta(z',z)
    \left[
        V^{\rm H,diss}_{in}(z)
        +V^{\rm Eh,diss}_{in}(z)
    \right]
    \nonumber\\
    &\quad
    +\Sigma^{\rm F,diss}_{in}(z',z).
    \label{eq:MF-diss-structure-main}
\end{align}
Thus, among the mean-field contributions,
only the dissipative Fock self-energy
explicitly connects opposite branches
of the Keldysh contour.

Combining the physical and dissipative
Ehrenfest contributions, we obtain
\begin{align}
    V^{\rm Eh}_{in}(z)
    &=V^{\rm Eh,phys}_{in}(z)
    +V^{\rm Eh,diss}_{in}(z)
    \nonumber\\
    &=\sum_\mu
    \mathcal G^{\rm el}_{\mu,in}(z)
    \langle\hat\varphi_\mu(z)\rangle,
    \label{eq:Eh-potential-main}
\end{align}
where $\mathcal G^{\rm el}$ is defined in
Eq.~\eqref{eq:effective-gp-main}.

Collecting the contour-local
mean-field contributions, we define
the self-adjoint coherent Hamiltonian
\begin{align}
    h^{\rm coh}_{in}(z)
    &=h^{(\Delta)}_{in}(z)
    +V^{\rm HF,phys}_{in}(z)
    +V^{\rm Eh}_{in}(z)
    +V^{\rm H,diss}_{in}(z).
    \label{eq:hcoh-definition-KBE-main}
\end{align}
It incorporates both physical and
dissipative renormalizations of
the coherent one-electron dynamics.

The dissipative Fock self-energy,
on the other hand, modifies the
gain and loss terms in the original
one-body kernel $\mathfrak h$.
To distinguish these additional
contributions from the single-electron
gain and loss potentials, we introduce
\begin{subequations}
\label{eq:fock-scattering-main}
\begin{align}
    W^<_{mn}(z)
    &\equiv
    \frac12\sum_{ij}
    v^{\rm diss}_{imjn}(z)
    \rho_{ji}(z),
    \\
    W^>_{mn}(z)
    &\equiv
    \frac12\sum_{ij}
    v^{\rm diss}_{minj}(z)
    [\delta_{ij}-\rho_{ji}(z)].
\end{align}
\end{subequations}
These matrices represent the additional
mean-field scattering generated by the
particle-bilinear jump operators.

Using the Fock expression in
Appendix~\ref{app:electron-KBE},
the dissipative Fock contribution
can be written as
\begin{align}
    \Sigma^{\rm F,diss}_{mn}(z',z)
    &=is(z)\delta(z',z)
    \left[
        V^{\rm diss}_{mn}(z)
        -W^>_{mn}(z)
        +W^<_{mn}(z)
    \right]
    \nonumber\\
    &\quad
    +2i\delta(z',z^*)
    \left[
        \theta_-(z)W^>_{mn}(z)
        +\theta_+(z)W^<_{mn}(z)
    \right].
    \label{eq:Fock-diss-contour-main}
\end{align}
Here $V^{\rm diss}$ is the one-body
potential defined in Eq.~\eqref{cpvdiss1}.
The term proportional to $V^{\rm diss}$
cancels the corresponding contribution
already present in $\mathfrak h$.
The remaining terms renormalize its
same-branch and opposite-branch
scattering structures.

The resulting lesser and greater
electronic scattering matrices are
\begin{subequations}
\label{eq:gamma-definition-KBE-main}
\begin{align}
    \gamma^<_{mn}(z)
    &=V^{\rm gain}_{mn}(z)
    +W^<_{mn}(z),
    \\
    \gamma^>_{mn}(z)
    &=V^{\rm loss}_{mn}(z)
    +W^>_{mn}(z),
    \\
    \gamma(z)
    &\equiv\gamma^<(z)+\gamma^>(z).
\end{align}
\end{subequations}
Here $V^{\rm gain}$ and $V^{\rm loss}$
are defined in Eq.~\eqref{vlossgain}.
The matrices $\gamma^<$ and $\gamma^>$
are self-adjoint and positive semidefinite
with respect to the electronic overlap
metric. They describe single-electron
gain and loss together with the
instantaneous dissipative scattering
generated by the particle-bilinear
jump operators.

Combining the local potentials and
the dissipative scattering contributions,
the complete mean-field self-energy
renormalizes the generalized one-body
$\mathfrak h$ of Eq.~\eqref{eq:hfrak-contour-main}
according to
\begin{align}
    \mathfrak h^{\rm MF}_{mn}(z',z)
    &=\mathfrak h_{mn}(z',z)
    +\Sigma^{\rm MF}_{mn}(z',z)
    \nonumber\\
    &=\delta(z',z)h^{\rm coh}_{mn}(z)
    \nonumber\\
    &\quad
    -is(z)\delta(z',z)
    \left[
        \gamma^>_{mn}(z)
        -\gamma^<_{mn}(z)
    \right]
    \nonumber\\
    &\quad
    +2i\delta(z',z^*)
    \left[
        \theta_-(z)\gamma^>_{mn}(z)
        +\theta_+(z)\gamma^<_{mn}(z)
    \right].
    \label{eq:hfrak-MF-contour-main}
\end{align}
The mean-field correction thus preserves
the contour structure of the original
one-body kernel. The physical Hartree-Fock,
Ehrenfest, and dissipative Hartree
contributions renormalize its local
self-adjoint part, whereas the dissipative
Fock contribution modifies the
scattering terms on the same and
opposite branches.

From the equal-time limit of the lesser
electronic KBEs, see Appendix~\ref{app:electron-KBE}, we get
\begin{align}
    \frac{d}{dt}\rho(t)
    &=-i[h^{\rm coh}(t),\rho(t)]
    -\{\gamma(t),\rho(t)\}
    \nonumber\\
    &\quad
    +2\gamma^<(t)
    +I_{\rm el}^{\rm corr}(t),
    \label{eq:rho-EOM-KBE-main}
\end{align}
where $\rho(t)=-iG^<(t,t)$
is the one-electron density matrix,
see Eq.~\eqref{defrho},
and the electronic correlation
collision integral
$I_{\rm el}^{\rm corr}$ is defined
in Eq.~\eqref{eq:electron-collision-app}.
The commutator describes coherent
evolution under the renormalized
Hamiltonian, whereas the anticommutator
and $2\gamma^<$ account for dissipative
scattering and its associated source
terms. The collision integral contains
correlation effects beyond mean field.

\subsection{Phoson covariance}
\label{subsec:equal-time-phoson}

For the phoson degrees of freedom, we introduce the
ordered connected covariance matrix
\begin{align}
    C^{\rm ph}_{\mu_1\mu_2}(t)
    &\equiv iD^<_{\mu_1\mu_2}(t,t)
    \nonumber\\
    &=\langle
        \hat\varphi_{\mu_2}(t)\hat\varphi_{\mu_1}(t)
      \rangle
      -\langle\hat\varphi_{\mu_2}(t)\rangle
       \langle\hat\varphi_{\mu_1}(t)\rangle.
    \label{eq:phoson-covariance-app}
\end{align}
Combining the left- and right-acting lesser phoson KBEs
in Eqs.~\eqref{eq:phoson-KBE-left-main} and
\eqref{eq:phoson-KBE-right-main} yields
\begin{align}
    \frac{d}{dt}C^{\rm ph}(t)
    &=-i\alpha\mathcal K(t)C^{\rm ph}(t)
     +iC^{\rm ph}(t)\mathcal K^{\mathsf T}(t)\alpha
    \nonumber\\
    &\quad+2\alpha\Lambda^<(t)\alpha
    +I_{\rm ph}^{\rm corr}(t),
    \label{eq:phoson-covariance-EOM-app}
\end{align}
where the phoson correlation collision integral is
\begin{align}
    I_{\rm ph}^{\rm corr}(t)
    &\equiv\left[
       \alpha I_{\rm ph}^{L}(t,t')
       -I_{\rm ph}^{R}(t,t')\alpha
      \right]_{t'=t}.
    \label{eq:phoson-collision-app}
\end{align}
The left and right correlation terms are
\begin{align}
    I_{\rm ph}^{L}(t,t')
    &\equiv\big[
        \Pi^{\rm R}\cdot D^<
        +\Pi^<\cdot D^{\rm A}
        +\Pi^{\rceil}\star D^{\lceil}
    \big](t,t'),
    \label{eq:phoson-collision-left-app}
    \\
    I_{\rm ph}^{R}(t,t')
    &\equiv\big[
        D^<\cdot\Pi^{\rm A}
        +D^{\rm R}\cdot\Pi^<
        +D^{\rceil}\star\Pi^{\lceil}
    \big](t,t').
    \label{eq:phoson-collision-right-app}
\end{align}
The collision integral includes electron-induced and
mixed dissipative correlations encoded in $\Pi$.
The explicit noise term $2\alpha\Lambda^<\alpha$
and the dissipative contribution to $\mathcal K$
originate from the same Lindblad jump operators.

The equal-time phoson commutation relations impose
\begin{align}
    [C^{\rm ph}(t)]^{\mathsf T}-C^{\rm ph}(t)
    &=\alpha,
    \label{eq:phoson-covariance-commutator-app}
\end{align}
whose preservation under the exact dynamics requires
\begin{align}
    [I_{\rm ph}^{\rm corr}(t)]^{\mathsf T}
    &=I_{\rm ph}^{\rm corr}(t).
    \label{eq:phoson-collision-symmetry-app}
\end{align}
This property is satisfied by the exact correlation
self-energy and provides a useful consistency condition
for approximate collision integrals.

\section{Stationarity-preserving jump operators}
\label{Stationarity-preserving jump operators}

We now illustrate the general formalism by
constructing jump operators that preserve a
reference state of a semiconductor with filled
valence bands, empty conduction bands, and
a phoson vacuum.
We identify dissipative channels compatible
with this reference state and verify its
stationarity using the equal-time equations [Eqs.\eqref{eq:rho-EOM-KBE-main} and \eqref{eq:phoson-covariance-EOM-app}]
of Sec.~\ref{sec:equal-time-KBE}.

We consider a crystalline semiconductor with
discrete translational symmetry and a finite
single-particle band gap, coupled to a set of
independent, undriven phoson normal modes.
These modes may represent either photons or
phonons, with annihilation operators
$\hat a_{\mathbf q\nu}$ and positive frequencies
$\omega_{\mathbf q\nu}$, where
$\nu=1,\ldots,N$.
Their quadratic Hamiltonian conserves phoson
number, and the linear source vanishes, $l=0$.
We choose a reference configuration in which
the filled valence bands exert no net mean
force on the phoson modes,
\begin{align}
    F^{\rm ph}_\mu[\rho_{\rm ref}]
    &=0,
    \label{eq:semiconductor-zero-force}
\end{align}
where $F^{\rm ph}$ is defined in
Eq.~\eqref{eq:phoson-force-main}.

For the electronic degrees of freedom,
we choose an orthonormal Bloch basis that
diagonalizes the self-consistent nondissipative
mean-field Hamiltonian
$h^{(\Delta)}+V^{\rm HF,phys}[\rho_{\rm ref}]$.
The corresponding occupied and unoccupied
eigenstates define the valence ($v$) and
conduction ($c$) bands, respectively.
The electronic states are labeled by
$(\mathbf k,n)$, where $\mathbf k$ is the
crystal momentum and $n$ includes the
band and spin indices.

The reference one-electron density matrix is
\begin{align}
    \rho_{{\rm ref},\mathbf k,n_1n_2}
    &=\delta_{n_1n_2}f_{\mathbf kn_1},
    \label{eq:semiconductor-rho-ref}
\end{align}
with
\begin{align}
    f_{\mathbf kv}
    &=\langle
        \hat d_{\mathbf kv}^\dagger
        \hat d_{\mathbf kv}
      \rangle_{\rm ref}=1,
    \nonumber\\
    f_{\mathbf kc}
    &=\langle
        \hat d_{\mathbf kc}^\dagger
        \hat d_{\mathbf kc}
      \rangle_{\rm ref}=0.
    \label{eq:semiconductor-occupations}
\end{align}
By construction, the reference density matrix
commutes with its self-consistent 
mean-field Hamiltonian,
\begin{align}
    [h^{(\Delta)}+V^{\rm HF,phys}[\rho_{\rm ref}],
     \rho_{\rm ref}]=0.
    \label{hcoh_com}
\end{align}

For the phoson degrees of freedom,
the reference state is the vacuum,
\begin{align}
    n_{\mathbf q\nu}
    &\equiv
    \langle
        \hat a_{\mathbf q\nu}^\dagger
        \hat a_{\mathbf q\nu}
    \rangle_{\rm ref}=0.
    \label{eq:semiconductor-phoson-occupation}
\end{align}
Since $l=0$, the displacement
$\Delta$ defined in
Eq.~\eqref{eq:Delta-contour-main}
vanishes, and $\hat\varphi=\hat\phi$.
Together with
Eq.~\eqref{eq:semiconductor-zero-force},
the one-point equation
\eqref{eq:shifted-Phi-EOM-main}
then admits the stationary solution
$\langle\hat\varphi\rangle_{\rm ref}=0$. Therefore $V^{\rm Eh}=0$.

In the following, we work within
the mean-field approximation,
$\Sigma^{\rm c}=\Pi=0$,
and construct a set of jump operators that preserve both the electronic
reference density matrix and the
phoson vacuum covariance.

We consider the following jump operators:
\begin{subequations}
\label{eq:semiconductor-jumps}
\begin{align}
    \hat L_{\mathbf kc}^{(1)}
    &=\eta^{\mathbf kc}
      \hat d_{\mathbf kc},
    \\
    \hat L_{\mathbf kv}^{(2)}
    &=\bar\eta^{\mathbf kv*}
      \hat d_{\mathbf kv}^\dagger,
    \\
    \hat L_{\mathbf q}^{(3)}
    &=\sum_{\mathbf kvc}
      \xi_{\mathbf kvc}^{\mathbf q}
      \hat d_{\mathbf k-\frac{\mathbf q}{2},v}^\dagger
      \hat d_{\mathbf k+\frac{\mathbf q}{2},c}+
      \sum_{\nu=1}^{N}
      \zeta_\nu^{\mathbf q}
      \hat a_{\mathbf q\nu}.
\end{align}
\end{subequations}
The first two operators describe
conduction-electron loss and valence-electron
gain, respectively.
The third combines interband electron-hole
recombination and phoson annihilation within
a common dissipative channel.
Each jump operator carries a definite
crystal-momentum transfer, preserving
the translational symmetry of the
Lindblad dynamics.

Using Eq.~\eqref{eq:v-diss-main},
the dissipative electron-electron interaction
induced by $\hat L_{\mathbf q}^{(3)}$ reads
\begin{align}
    &v^{\rm diss}_{
    \mathbf k_1n_1\mathbf k_2n_2
    \mathbf k_3n_3\mathbf k_4n_4}
    \nonumber\\
    &=2\delta_{
      \mathbf k_1-\mathbf k_4,
      \mathbf k_3-\mathbf k_2}
    \nonumber\\
    &\quad\times
    \xi_{
      \frac{\mathbf k_1+\mathbf k_4}{2},
      n_4n_1}^{
      \mathbf k_1-\mathbf k_4 *}
    \xi_{
      \frac{\mathbf k_2+\mathbf k_3}{2},
      n_2n_3}^{
      \mathbf k_3-\mathbf k_2}
    F_{n_1n_2n_3n_4}.
    \label{eq:semiconductor-vdiss}
\end{align}
Here $F_{cvc'v'}=1$ for conduction
indices $c,c'$ and valence indices
$v,v'$, and vanishes for all other
band combinations.
The momentum-conserving Kronecker delta
reflects the translational symmetry
of the dissipative interaction.

We first verify the stationarity of
the electronic reference density matrix.
For the interband recombination jumps,
the dissipative Hartree potential
$V^{\rm H,diss}$ of Eq.~\eqref{vadiss}
vanishes when evaluated with
$\rho_{\rm ref}$.
The dissipative Fock contributions $W^\lessgtr$ to
the scattering matrices
$\gamma^{\lessgtr}$ in
Eq.~\eqref{eq:gamma-definition-KBE-main}
also vanish for this reference state.
Consequently, the scattering matrices
are diagonal in the band indices,
with
\begin{subequations}
\label{gamma><semi}
\begin{align}
    \gamma_{\mathbf kn}^{>}
    &=\delta_{n\{c\}}
      |\eta^{\mathbf kn}|^2,
    \\
    \gamma_{\mathbf kn}^{<}
    &=\delta_{n\{v\}}
      |\bar\eta^{\mathbf kn}|^2.
\end{align}
\end{subequations}
Here $\delta_{n\{c\}}$ and
$\delta_{n\{v\}}$ equal unity when
$n$ belongs to the conduction and
valence bands, respectively, and
vanish otherwise.

Since $V^{\rm Eh}$ and $V^{\rm H,diss}$
vanish for the reference state,
Eq.~\eqref{hcoh_com} also implies
$[h^{\rm coh},\rho_{\rm ref}]=0$.
The stationary electronic mean-field
equation follows from
Eq.~\eqref{eq:rho-EOM-KBE-main},
\begin{align}
    0
    &=-i[h^{\rm coh},\rho]
      -\{\gamma,\rho\}
      +2\gamma^<.
    \label{eq:semiconductor-electron-stationary}
\end{align}
Using Eq.~\eqref{hcoh_com}
and the diagonal structure of
$\rho_{\rm ref}$ and
$\gamma^{\lessgtr}$, we obtain
\begin{align}
    \gamma_{\mathbf kn}
    \rho_{\mathbf k,nn}
    &=\gamma_{\mathbf kn}^{<},
    \label{eq:semiconductor-rate-equation}
\end{align}
where
$\gamma_{\mathbf kn}
=\gamma_{\mathbf kn}^{>}
+\gamma_{\mathbf kn}^{<}$.
For $\gamma_{\mathbf kn}>0$,
the stationary solution is
\begin{align}
    \rho_{\mathbf k,n_1n_2}
    &=\delta_{n_1n_2}
    \frac{\gamma_{\mathbf kn_1}^{<}}
    {\gamma_{\mathbf kn_1}^{>}
     +\gamma_{\mathbf kn_1}^{<}}
    \nonumber\\
    &=\delta_{n_1n_2}
      f_{\mathbf kn_1}.
    \label{eq:semiconductor-rho-stationary}
\end{align}
For modes with vanishing total decay rate, the stationary
equation imposes no additional constraint on their occupations;
the reference occupations nevertheless remain
stationary under the assumed coherent dynamics.

Thus, the filled-valence and
empty-conduction configuration
is a stationary solution of the
electronic mean-field KBE.

We now verify the stationarity of
the phoson vacuum.
For each crystal momentum $\mathbf q$,
we introduce the column vector
\begin{align}
    \hat{\bm a}_{\mathbf q}
    &=
    (\hat a_{\mathbf q1},\ldots,
     \hat a_{\mathbf qN})^{\mathsf T},
    \label{eq:semiconductor-phoson-vector}
\end{align}
and the corresponding doubled-field
representation
\begin{align}
    \hat\phi_{\mathbf q}
    &=
    \begin{pmatrix}
        \hat{\bm a}_{\mathbf q}\\
        (\hat{\bm a}_{\mathbf q}^{\dagger})^{\mathsf T}
    \end{pmatrix}.
    \label{eq:semiconductor-phoson-field}
\end{align}
The commutation matrix and quadratic
Hamiltonian coefficient are
\begin{align}
    \alpha_{\mathbf q}
    &=
    \begin{pmatrix}
        0&I\\
        -I&0
    \end{pmatrix},
    &
    Q_{\mathbf q}
    &=
    \begin{pmatrix}
        0&\Omega_{\mathbf q}\\
        \Omega_{\mathbf q}&0
    \end{pmatrix},
    \label{eq:semiconductor-phoson-matrices}
\end{align}
where
\begin{align}
    \Omega_{\mathbf q}
    &=
    \operatorname{diag}
    (\omega_{\mathbf q1},\ldots,
     \omega_{\mathbf qN}).
    \label{eq:semiconductor-omega}
\end{align}

Recall from Eq.~\eqref{eq:K-main}
that the effective phoson kernel is
$\mathcal K=Q+i(\Lambda^<-\Lambda^>)$.
The matrices $\Lambda^\lessgtr$
determine both the dissipative
contribution to $\mathcal K$ and
the noise term in the equal-time
phoson KBE,
Eq.~\eqref{eq:phoson-covariance-EOM-app}.

We introduce the Hermitian
positive-semidefinite damping matrix
\begin{align}
    (B_{\mathbf q})_{\nu\nu'}
    &\equiv
    \zeta_\nu^{\mathbf q*}
    \zeta_{\nu'}^{\mathbf q}.
    \label{eq:semiconductor-B}
\end{align}
For the jumps in
Eq.~\eqref{eq:semiconductor-jumps},
the dissipative matrices are
\begin{subequations}
\label{eq:semiconductor-Lambda}
\begin{align}
    \Lambda_{\mathbf q}^<
    &=
    \begin{pmatrix}
        0&B_{\mathbf q}^{\mathsf T}\\
        0&0
    \end{pmatrix},
    \\
    \Lambda_{\mathbf q}^>
    &=(\Lambda_{\mathbf q}^<)^{\mathsf T}.
\end{align}
\end{subequations}
The effective quadratic phoson
kernel therefore reads
\begin{align}
    \mathcal K_{\mathbf q}
    &=Q_{\mathbf q}
    +i(\Lambda_{\mathbf q}^<
       -\Lambda_{\mathbf q}^>)
    \nonumber\\
    &=
    \begin{pmatrix}
        0&
        \Omega_{\mathbf q}
        +iB_{\mathbf q}^{\mathsf T}\\
        \Omega_{\mathbf q}
        -iB_{\mathbf q}&0
    \end{pmatrix}.
    \label{eq:semiconductor-K}
\end{align}

The connected phoson covariance
corresponding to the vacuum is
\begin{align}
    C_{\mathbf q}^{\rm ph}
    &=
    \begin{pmatrix}
        0&0\\
        I&0
    \end{pmatrix}.
    \label{eq:semiconductor-Cvac}
\end{align}
It satisfies the canonical
commutation constraint
\begin{align}
    (C_{\mathbf q}^{\rm ph})^{\mathsf T}
    -C_{\mathbf q}^{\rm ph}
    &=\alpha_{\mathbf q}.
    \label{eq:semiconductor-C-constraint}
\end{align}

With $\Pi=0$, the equal-time phoson
KBE of Eq.~\eqref{eq:phoson-covariance-EOM-app}
reduces to
\begin{align}
    \frac{d}{dt}C_{\mathbf q}^{\rm ph}
    &=-i\alpha_{\mathbf q}
      \mathcal K_{\mathbf q}
      C_{\mathbf q}^{\rm ph}+
      iC_{\mathbf q}^{\rm ph}
      \mathcal K_{\mathbf q}^{\mathsf T}
      \alpha_{\mathbf q}
    +2\alpha_{\mathbf q}
      \Lambda_{\mathbf q}^{<}
      \alpha_{\mathbf q}.
    \label{eq:semiconductor-C-EOM}
\end{align}
Substituting the vacuum covariance
of Eq.~\eqref{eq:semiconductor-Cvac},
the drift terms give
\begin{align}
    &-i\alpha_{\mathbf q}
      \mathcal K_{\mathbf q}
      C_{\mathbf q}^{\rm ph}
    +iC_{\mathbf q}^{\rm ph}
      \mathcal K_{\mathbf q}^{\mathsf T}
      \alpha_{\mathbf q}=
    \begin{pmatrix}
        0&0\\
        -2B_{\mathbf q}^{\mathsf T}&0
    \end{pmatrix},
    \label{eq:semiconductor-drift}
\end{align}
while the noise term is
\begin{align}
    2\alpha_{\mathbf q}
    \Lambda_{\mathbf q}^{<}
    \alpha_{\mathbf q}
    &=
    \begin{pmatrix}
        0&0\\
        2B_{\mathbf q}^{\mathsf T}&0
    \end{pmatrix}.
    \label{eq:semiconductor-noise}
\end{align}
The drift and noise contributions
cancel exactly, yielding
\begin{align}
    \frac{d}{dt}C_{\mathbf q}^{\rm ph}
    &=0.
    \label{eq:semiconductor-C-stationary}
\end{align}
Therefore, the phoson vacuum
covariance is stationary under
the dissipative mean-field dynamics.
The phoson one-point function
also remains zero under the
conditions of
Eq.~\eqref{eq:semiconductor-zero-force}.

The two-time Green’s functions at the same mean-field level are derived in Appendix~\ref{app:mf-KBE-solution}.

\section{Conclusions}
\label{sec:conclusions}

We have developed a dissipative nonequilibrium Green's function theory
for interacting electrons coupled to photon and phonon fields. By
combining the Keldysh-Lindblad formulation with an explicit treatment
of bosonic degrees of freedom, the approach extends dissipative many-body
diagrammatics to Hamiltonians that need not conserve boson number. It also
allows electronic transitions and bosonic processes to share a dissipative
channel, so that the environment can couple the two sectors rather than
simply broaden their excitations independently.

The resulting electron and phoson KBEs describe
nonequilibrium propagation together with quantum jumps, environmental
noise, and interaction-induced memory. Their equal-time limits provide
evolution equations for the electronic density matrix and bosonic
covariance. As a concrete illustration, we identified jump operators
that preserve the filled-valence and phoson-vacuum configuration of
a semiconductor. Within the mean-field approximation, the one-body
electronic density matrix and phoson covariance remain stationary,
while the corresponding two-time correlations acquire finite lifetimes.

An important advantage of the present formalism is its ability to
describe the real-time evolution of two-time electron and phoson
Green's functions, beyond the equal-time observables commonly
considered in Lindblad dynamics. These correlation functions
contain information about the spectral properties and lifetimes
of excitations that cannot be inferred from the density matrices
alone. In particular, the electronic lesser Green's function
$G^<(t,t')$ provides the information needed to calculate
time-resolved photoemission spectra
\cite{Freericks2009,gianluca2026exciton-prl,Caruso2020}.
The phonon components of $D^{\lessgtr}(t,t')$
provide vibrational correlation functions
relevant to time-resolved Raman spectroscopy,
together with the appropriate Raman coupling
vertices~\cite{Okumura1997Raman,Versteeg2018}.

Similarly, the photon components provide the field
correlations underlying photoluminescence spectra
\cite{Ishida2013}.
Thus, the formalism offers a direct connection between
dissipative many-body dynamics and experimentally
accessible time-dependent spectra.

While two-time correlations can also be obtained from
Lindblad master equations using the quantum regression
theorem~\cite{Lax1963}, the present Green's function formulation provides
a systematic diagrammatic approach to interacting many-body
systems. It allows electronic and bosonic propagators to
be dressed self-consistently and incorporates interaction-induced
memory and initial correlations within a unified framework.
This provides a basis for developing controlled many-body
approximations and investigating the dynamics and spectra
of quantum materials in lossy photonic and vibrational
environments.

\acknowledgments

We acknowledge funding from Ministero Universit\`a e 
Ricerca PRIN under grant agreement No. 2022WZ8LME, 
from INFN through project TIME2QUEST, 
from European Research Council MSCA-ITN TIMES under grant agreement 101118915, 
and from Tor Vergata University through project TESLA.

\appendix

\section{Electron Kadanoff-Baym equations}
\label{app:electron-KBE}

For completeness, we present the electronic
Hartree and Fock self-energies and the real-time,
mixed, and Matsubara KBEs.
The conventions for contour components and
time convolutions are given in
Sec.~\ref{sec:phoson-KBE}.
The coherent Hamiltonian $h^{\rm coh}$ and
the scattering matrices $\gamma^{\lessgtr}$ are defined
in Sec.~\ref{subsec:equal-time-electron}.
For the two-time electronic KBEs,
it is convenient to introduce the
non-self-adjoint mean-field Hamiltonian
\begin{subequations}
\label{eq:open-h-main}
\begin{align}
    h^{\rm MF}(z)
    &\equiv h^{\rm coh}(z)-i\gamma(z),
    \\
    h^{\rm MF,ad}(z)
    &=h^{\rm coh}(z)+i\gamma(z)
    \nonumber\\
    &=\mathcal M^{-1}
    [h^{\rm MF}(z)]^\dagger\mathcal M.
\end{align}
\end{subequations}

As discussed in Eq.~\eqref{eq:electron-selfenergy-split-main},
the electronic skeleton self-energy contains
the Hartree, Fock, and correlation contributions.
The Hartree and Fock terms are
\begin{subequations}
\label{eq:HF-diagrams}
\begin{align}
    \Sigma^{\rm H}_{in}(z,z')
    &=-i\delta(z,z')
    \sum_{jm}\int_C d\bar z\,
    v_{ijmn}(\bar z,z)
    G_{mj}(\bar z,\bar z^+)
    \nonumber\\
    &=\delta(z,z')
    \sum_{jm}
    \left[
        v^{\rm phys}_{ijmn}(z)
        -iv^{\rm a,diss}_{ijmn}(z)
    \right]\rho_{mj}(z),
    \label{eq:electron-Hartree-app}
    \\
    \Sigma^{\rm F}_{in}(z,z')
    &=i\sum_{jm}
    v_{ijnm}(z',z)G_{mj}(z,z').
    \label{eq:electron-Fock-app}
\end{align}
\end{subequations}
Here $v(z,z')$ is the generalized interaction
kernel defined in Eq.~\eqref{eq:v-contour}.
The resulting local potentials and dissipative
scattering matrices are derived in
Sec.~\ref{subsec:equal-time-electron}.
The Ehrenfest one-point contribution is
incorporated separately through the potential
$V^{\rm Eh}$ of Eq.~\eqref{eq:Eh-potential-main},
and is not included in the skeleton self-energy.

Combining the mean-field contributions with
the generalized one-body kernel, as in
Eq.~\eqref{eq:hfrak-MF-contour-main}, the
electronic contour equation takes the form
\begin{align}
    i\frac{d}{dz}G(z,z')
    &-\int_C d\bar z\,
    \left[
        \mathfrak h^{\rm MF}(z,\bar z)
        +\Sigma^{\rm c}(z,\bar z)
    \right]G(\bar z,z')
    \nonumber\\
    &=\delta(z,z')I.
    \label{eom1}
\end{align}

Since the jump amplitudes vanish on the
Matsubara branch, the scattering matrices
satisfy $\gamma^{\lessgtr}(t_0-i\tau)=0$.
Consequently, the mean-field Hamiltonian
on this branch is self-adjoint,
\begin{align}
    h^{{\rm MF},{\rm M}}
    &\equiv h^{\rm MF}(t_0-i\tau)
    \nonumber\\
    &=h^{\rm MF,ad}(t_0-i\tau)
    =h^{\rm coh}(t_0-i\tau).
    \label{eq:electron-MF-Matsubara-app}
\end{align}
It is independent of $\tau$ for the
stationary Matsubara initial state.

Applying the Langreth rules
~\cite{Langreth1976,StefanucciVanLeeuwen2025}
to Eq.~\eqref{eom1}, we obtain the
left- and right-acting electronic
lesser and greater KBEs
~\cite{stefanucci2024kadanoff,blommel_unified_2025},
\begin{subequations}
\label{eq:electron-KBE}
\begin{align}
    &\left[
        i\frac{d}{dt}-h^{\rm MF}(t)
    \right]G^\lessgtr(t,t')
    \mp2i\gamma^\lessgtr(t)G^{\rm A}(t,t')
    \nonumber\\
    &\qquad=
    \big[
        \Sigma^{\rm c,R}\cdot G^\lessgtr
        +\Sigma^{\rm c,\lessgtr}\cdot G^{\rm A}
        +\Sigma^{\rm c,\rceil}\star G^{\lceil}
    \big](t,t'),
    \label{eq:electron-KBE-left-main}
    \\
    &G^\lessgtr(t,t')
    \left[
        -i\frac{\overleftarrow d}{dt'}
        -h^{\rm MF,ad}(t')
    \right]
    \mp2iG^{\rm R}(t,t')\gamma^\lessgtr(t')
    \nonumber\\
    &\qquad=
    \big[
        G^\lessgtr\cdot\Sigma^{\rm c,A}
        +G^{\rm R}\cdot\Sigma^{\rm c,\lessgtr}
        +G^{\rceil}\star\Sigma^{\rm c,\lceil}
    \big](t,t').
    \label{eq:electron-KBE-right-main}
\end{align}
\end{subequations}
The upper and lower signs correspond
to the lesser and greater components,
respectively.

The retarded electronic Green's function obeys
\begin{align}
    \left[
        i\frac{d}{dt}-h^{\rm MF}(t)
    \right]G^{\rm R}(t,t')
    &=\delta(t-t')I
    \nonumber\\
    &\quad+
    [\Sigma^{\rm c,R}\cdot G^{\rm R}](t,t').
    \label{eq:electron-retarded-app}
\end{align}
The advanced component follows from
$G^{\rm A}(t,t')
=[G^{\rm R}(t',t)]^{\rm ad}$.

The right-mixing component satisfies
\begin{subequations}
\begin{align}
    \left[
        i\frac{d}{dt}-h^{\rm MF}(t)
    \right]G^{\rceil}(t,\tau)
    &=
    \big[
        \Sigma^{\rm c,R}\cdot G^{\rceil}
        +\Sigma^{\rm c,\rceil}\star G^{\rm M}
    \big](t,\tau),
    \label{eq:electron-rceil-real-app}
    \\
    G^{\rceil}(t,\tau)
    \left[
        \frac{\overleftarrow d}{d\tau}
        -h^{{\rm MF},{\rm M}}
    \right]
    &=
    \big[
        G^{\rm R}\cdot\Sigma^{\rm c,\rceil}
        +G^{\rceil}\star\Sigma^{\rm c,M}
    \big](t,\tau).
    \label{eq:electron-rceil-M-app}
\end{align}
\end{subequations}

Similarly, the left-mixing component obeys
\begin{subequations}
\begin{align}
    \left[
        -\frac{d}{d\tau}
        -h^{{\rm MF},{\rm M}}
    \right]G^{\lceil}(\tau,t)
    &=
    \big[
        \Sigma^{\rm c,M}\star G^{\lceil}
        +\Sigma^{\rm c,\lceil}\cdot G^{\rm A}
    \big](\tau,t),
    \label{eq:electron-lceil-M-app}
    \\
    G^{\lceil}(\tau,t)
    \left[
        -i\frac{\overleftarrow d}{dt}
        -h^{\rm MF,ad}(t)
    \right]
    &=
    \big[
        G^{\lceil}\cdot\Sigma^{\rm c,A}
        +G^{\rm M}\star\Sigma^{\rm c,\lceil}
    \big](\tau,t).
    \label{eq:electron-lceil-real-app}
\end{align}
\end{subequations}

On the Matsubara branch, the electronic
Green's function satisfies
\begin{subequations}
\begin{align}
    \left[
        -\frac{d}{d\tau}
        -h^{{\rm MF},{\rm M}}
    \right]G^{\rm M}(\tau,\tau')
    &=i\delta(\tau-\tau')I
    \nonumber\\
    &\quad+
    [\Sigma^{\rm c,M}\star G^{\rm M}]
    (\tau,\tau'),
    \label{eq:electron-M-left-app}
    \\
    G^{\rm M}(\tau,\tau')
    \left[
        \frac{\overleftarrow d}{d\tau'}
        -h^{{\rm MF},{\rm M}}
    \right]
    &=i\delta(\tau-\tau')I
    \nonumber\\
    &\quad+
    [G^{\rm M}\star\Sigma^{\rm c,M}]
    (\tau,\tau').
    \label{eq:electron-M-right-app}
\end{align}
\end{subequations}

The equal-time limit of the lesser
electronic KBEs yields the density-matrix
equation~\eqref{eq:rho-EOM-KBE-main},
where the correlation collision integral is
\begin{align}
    I_{\rm el}^{\rm corr}(t)
    &\equiv
    -\left[
        I_{\rm el}^{L}(t,t')
        -I_{\rm el}^{R}(t,t')
    \right]_{t'=t}.
    \label{eq:electron-collision-app}
\end{align}
The left and right correlation terms are
\begin{align}
    I_{\rm el}^{L}(t,t')
    &\equiv
    \big[
        \Sigma^{\rm c,R}\cdot G^<
        +\Sigma^{\rm c,<}\cdot G^{\rm A}
        +\Sigma^{\rm c,\rceil}\star G^{\lceil}
    \big](t,t'),
    \label{eq:electron-collision-left-app}
    \\
    I_{\rm el}^{R}(t,t')
    &\equiv
    \big[
        G^<\cdot\Sigma^{\rm c,A}
        +G^{\rm R}\cdot\Sigma^{\rm c,<}
        +G^{\rceil}\star\Sigma^{\rm c,\lceil}
    \big](t,t').
    \label{eq:electron-collision-right-app}
\end{align}

Taking the one-electron trace of
Eq.~\eqref{eq:rho-EOM-KBE-main}
gives the particle-number balance,
\begin{align}
    \frac{d}{dt}\Tr\rho(t)
    &=-2\Tr[V^{\rm loss}(t)\rho(t)]
    \nonumber\\
    &\quad+
    2\Tr[V^{\rm gain}(t)(I-\rho(t))]
    \nonumber\\
    &\quad+
    \Tr I_{\rm el}^{\rm corr}(t).
    \label{eq:electron-number-balance-KBE-main}
\end{align}
The contributions from the
electron-number-conserving jump operators
cancel in the trace.
For the exact correlation self-energy,
\begin{align}
    \Tr I_{\rm el}^{\rm corr}(t)
    &=0,
    \label{eq:electron-collision-number-app}
\end{align}
consistent with the global $U(1)$
symmetry of the Hamiltonian and
particle-bilinear jump operators.
Consequently, the total electron number
changes only through the single-electron
loss and gain channels.
Conserving approximations must preserve
the same collision identity
~\cite{zhang2026nhbse}.

\section{Analytical Mean-Field Green's Functions
for the Stationary Semiconductor}
\label{app:mf-KBE-solution}
In this appendix, we derive the analytical two-time
Green's functions for the stationary semiconductor
model introduced in Sec.~\ref{Stationarity-preserving jump operators}. Within the mean-field
approximation, we consider time-independent coefficients
for $t\geq t_0$ and the filled-valence electronic
configuration with the phoson vacuum.

\subsection{Real-time electronic Green's functions}

In the band-diagonal representation, the
electronic mean-field Hamiltonian in Eq.~\eqref{eq:open-h-main} reads
\begin{align}
    h^{\rm MF}_{\mathbf k,n_1n_2}
    &=\delta_{n_1n_2}E_{\mathbf kn_1},
\end{align}
where
\begin{align}
    E_{\mathbf kn}
    &\equiv
    \varepsilon_{\mathbf kn}
    -i\gamma_{\mathbf kn}.
\end{align}
Here $\varepsilon_{\mathbf kn}$ denotes the
self-consistent mean-field energy, while
$\gamma_{\mathbf kn}\geq0$ is the corresponding
decay rate.

For convenience, we define the nonunitary
single-particle propagator
\begin{align}
    u_{\mathbf kn}(s)
    &\equiv e^{-iE_{\mathbf kn}s},
    \qquad s\geq0.
\end{align}
The retarded and advanced electronic
Green's functions are then
\begin{align}
    G_{\mathbf kn}^{\rm R}(t,t')
    &=-i\theta(t-t')
    u_{\mathbf kn}(t-t'),
    \\
    G_{\mathbf kn}^{\rm A}(t,t')
    &=i\theta(t'-t)
    u_{\mathbf kn}^*(t'-t).
    \label{eq:semiconductor-GRA}
\end{align}
These propagators describe damped electron
and hole excitations, with the retarded
and advanced components related by
$G^{\rm A}(t,t')
=[G^{\rm R}(t',t)]^\dagger$.

For the stationary occupations
$f_{\mathbf kn}$, the lesser and greater
Green's functions take the form
\begin{align}
    G_{\mathbf kn}^{<}(t,t')
    &=if_{\mathbf kn}
    e^{-i\varepsilon_{\mathbf kn}(t-t')}
    e^{-\gamma_{\mathbf kn}|t-t'|},
    \\
    G_{\mathbf kn}^{>}(t,t')
    &=-i(1-f_{\mathbf kn})
    e^{-i\varepsilon_{\mathbf kn}(t-t')}
    e^{-\gamma_{\mathbf kn}|t-t'|}.
    \label{eq:semiconductor-Glessgreater}
\end{align}
The absolute value in the damping factor
arises from the two time orderings of the
left- and right-acting KBEs.
For $t>t'$, the propagation is governed by
$h^{\rm MF}$, whereas for $t<t'$ it is
governed by $h^{\rm MF,ad}$.
Both time orderings therefore yield
exponential decay with increasing
$|t-t'|$.

The equal-time limits reproduce the
stationary occupations,
\begin{align}
    -iG_{\mathbf kn}^{<}(t,t)
    &=f_{\mathbf kn},
    \\
    iG_{\mathbf kn}^{>}(t,t)
    &=1-f_{\mathbf kn}.
\end{align}
Thus, the electronic populations remain
stationary even though the two-time
correlations acquire finite dissipative
linewidths.

\subsection{Real-time phoson Green's functions}

For the phoson sector, we introduce the
nonunitary annihilation-mode propagator
\begin{align}
    U_{\mathbf q}(s)
    &\equiv
    \exp\left[
        (-i\Omega_{\mathbf q}
        -B_{\mathbf q})s
    \right],
    \qquad s\geq0.
    \label{eq:semiconductor-Uph}
\end{align}
Its doubled-field counterpart is
\begin{align}
    \mathcal U_{\mathbf q}(s)
    &\equiv
    \begin{pmatrix}
        U_{\mathbf q}(s)&0\\
        0&U_{\mathbf q}^*(s)
    \end{pmatrix}
    \nonumber\\
    &=e^{-i\alpha_{\mathbf q}
    \mathcal K_{\mathbf q}s}.
\end{align}
The retarded phoson Green's function is
\begin{align}
    D_{\mathbf q}^{\rm R}(t,t')
    &=-i\theta(t-t')
    \mathcal U_{\mathbf q}(t-t')
    \alpha_{\mathbf q},
    \label{eq:semiconductor-DR}
\end{align}
while the advanced component follows from
\begin{align}
    D_{\mathbf q}^{\rm A}(t,t')
    &=
    [D_{\mathbf q}^{\rm R}(t',t)]^{\mathsf T}.
    \label{eq:semiconductor-DA}
\end{align}

To express the lesser and greater components
compactly, we define the two-time propagator
\begin{align}
    \mathscr U_{\mathbf q}(t,t')
    &\equiv
    \begin{cases}
        U_{\mathbf q}(t-t'),
        &t\geq t',\\
        U_{\mathbf q}^\dagger(t'-t),
        &t<t'.
    \end{cases}
    \label{eq:semiconductor-two-time-U}
\end{align}
For the phoson vacuum, only contractions
between phoson annihilation and creation
operators survive.
The lesser and greater Green's functions
therefore read
\begin{align}
    D_{\mathbf q}^{>}(t,t')
    &=-i
    \begin{pmatrix}
        0&\mathscr U_{\mathbf q}(t,t')\\
        0&0
    \end{pmatrix},
    \\
    D_{\mathbf q}^{<}(t,t')
    &=-i
    \begin{pmatrix}
        0&0\\
        \mathscr U_{\mathbf q}^{\mathsf T}(t',t)&0
    \end{pmatrix}.
    \label{eq:semiconductor-Dlessgreater}
\end{align}
They satisfy
\begin{align}
    D_{\mathbf q}^{\rm R}(t,t')
    &=\theta(t-t')
    \left[
        D_{\mathbf q}^{>}(t,t')
        -D_{\mathbf q}^{<}(t,t')
    \right],
\end{align}
together with the corresponding relation
for $D_{\mathbf q}^{\rm A}$.

In particular, their equal-time limit gives
\begin{align}
    iD_{\mathbf q}^{<}(t,t)
    &=
    \begin{pmatrix}
        0&0\\
        I&0
    \end{pmatrix}
    =C_{\mathbf q}^{\rm ph},
\end{align}
consistent with the stationary phoson
covariance obtained above.

\subsection{Matsubara and mixed components}

We choose the zero-temperature Matsubara
reference state to coincide with the
filled-valence electronic configuration
and the phoson vacuum.
Accordingly, we take the limit
$\beta\rightarrow\infty$, with the chemical
potential $\mu$ lying inside the band gap.

For the electronic Matsubara Hamiltonian,
we define
\begin{align}
    \varepsilon_{\mathbf kn}^{\rm M}
    &\equiv
    \varepsilon_{\mathbf kn}-\mu,
\end{align}
where
$\varepsilon_{\mathbf kv}^{\rm M}<0$ and
$\varepsilon_{\mathbf kc}^{\rm M}>0$.
The electronic Matsubara Green's function is
\begin{align}
    G_{\mathbf kn}^{\rm M}(\tau,\tau')
    &=-i
    e^{-\varepsilon_{\mathbf kn}^{\rm M}(\tau-\tau')}
    \nonumber\\
    &\quad\times
    \left[
        (1-f_{\mathbf kn})\theta(\tau-\tau')
        -f_{\mathbf kn}\theta(\tau'-\tau)
    \right].
    \label{eq:semiconductor-GM}
\end{align}
This expression satisfies the fermionic
Matsubara equations and the zero-temperature
limit of the KMS boundary conditions.

The right- and left-mixing electronic
Green's functions are
\begin{align}
    G_{\mathbf kn}^{\rceil}(t,\tau)
    &=if_{\mathbf kn}
    u_{\mathbf kn}(t-t_0)
    e^{\varepsilon_{\mathbf kn}^{\rm M}\tau},
    \\
    G_{\mathbf kn}^{\lceil}(\tau,t)
    &=-i(1-f_{\mathbf kn})
    e^{-\varepsilon_{\mathbf kn}^{\rm M}\tau}
    u_{\mathbf kn}^*(t-t_0).
    \label{eq:semiconductor-Gmixed}
\end{align}
These expressions combine the imaginary-time
evolution governed by the self-adjoint
Matsubara Hamiltonian with the dissipative
real-time propagation generated by
$h^{\rm MF}$ and $h^{\rm MF,ad}$.

For the phoson Matsubara reference problem,
all jump amplitudes vanish, and hence
$\mathcal K_{\mathbf q}^{\rm M}
=Q_{\mathbf q}$.
The phoson
Matsubara Green's function reads
\begin{align}
    D_{\mathbf q}^{\rm M}(\tau,\tau')
    &=-i\theta(\tau-\tau')
    \begin{pmatrix}
        0&e^{-\Omega_{\mathbf q}(\tau-\tau')}\\
        0&0
    \end{pmatrix}
    \nonumber\\
    &\quad
    -i\theta(\tau'-\tau)
    \begin{pmatrix}
        0&0\\
        e^{\Omega_{\mathbf q}(\tau-\tau')}&0
    \end{pmatrix}.
    \label{eq:semiconductor-DM}
\end{align}
It satisfies the bosonic Matsubara equations
with the vacuum boundary conditions.

The corresponding phoson mixed components are
\begin{align}
    D_{\mathbf q}^{\rceil}(t,\tau)
    &=-i
    \begin{pmatrix}
        0&0\\
        U_{\mathbf q}^*(t-t_0)
        e^{-\Omega_{\mathbf q}\tau}&0
    \end{pmatrix},
    \\
    D_{\mathbf q}^{\lceil}(\tau,t)
    &=-i
    \begin{pmatrix}
        0&e^{-\Omega_{\mathbf q}\tau}
        U_{\mathbf q}^\dagger(t-t_0)\\
        0&0
    \end{pmatrix}.
    \label{eq:semiconductor-Dmixed}
\end{align}
As in the electronic case, the Matsubara
time dependence is governed by the closed
reference Hamiltonian, while the real-time
dependence incorporates the dissipative
phoson propagation.

We have thus obtained an analytic mean-field
solution in which the filled-valence electronic
configuration and phoson vacuum remain stationary,
while the corresponding two-time Green's functions
exhibit dissipative broadening.
This solution provides a reference for studying
electron-phoson correlations beyond the
mean-field approximation.

\bibliography{references}

\end{document}